\documentclass{article}

\usepackage{silence}
\usepackage{tikz}                               % creating graphics and diagrams
\usetikzlibrary{arrows,through,calc,shapes}     % TikZ library for arrow types and geometric constructions
\usepackage{geometry}                           % adjusting the page layout
\usepackage{booktabs}                           % creating tables with improved rules
\usepackage{array}                              % enhancing column formatting in tables
\usepackage{lmodern}                            % improved fonts (incl. better \texteuro)
\usepackage{textcomp}                           % \texteuro and other text symbols
\usepackage[gen]{eurosym}                       % defines \EUR{...}
\usepackage{colortbl}                           % adding color to tables
\usepackage{subcaption}                         % creating subfigures within a figure
\usepackage{amsmath}                            % supporting mathematics
\usepackage{amssymb}                            % for \mathbb and additional symbols
\usepackage{float}                              % managing the position of figures
\usepackage{nicematrix}                         % matrix with row and column indexes
\usepackage[hidelinks]{hyperref}                % remove ugly rectangles around hyperlinks
\usepackage{orcidlink}                          % use ORCID
\usepackage{tabularx}                           % package for creating tables with variable width columns
\usepackage{microtype}                          % improves text appearance with micro-typographic adjustments
\usepackage{newunicodechar} % define behavior for specific Unicode characters (XeLaTeX/LuaLaTeX)

\renewcommand{\EUR}[1]{\texteuro{}#1}
\newunicodechar{€}{\texteuro}

\title{Deepening the Secondary Market: \\ Integrating Trade Credit into Market Clearing with the Cycles Protocol}
\author{Tomaž Fleischman\,\orcidlink{0000-0002-9767-8584}, Ethan Buchman \\ \small Cycles Protocol SA}
\date{March 2026}

\begin{document}

\maketitle

\begin{abstract}

Current post-trade clearing systems rely almost exclusively on cash or cash-like collateral, 
leaving vast reserves of short-term liquidity embedded in trade credit outside formal settlement infrastructures. A key barrier to integrating this liquidity is the near-universal dependence of clearing services on novation, which imposes institutional overhead that restricts accessibility and limits the range of obligations that can be brought into settlement.

This paper introduces the Cycles Protocol: a distributed, multilateral clearing mechanism based on double-entry accounting and atomic cycle execution that maximizes balance sheet compression. Unlike novation-based clearing, Cycles does not redistribute counterparty risk; it can thus be applied generally to existing financial networks, without any change in counterparty relations, allowing it to complement existing clearing systems and Central Counterparties (CCPs).

By representing commitments as edges on a unified directed graph, Cycles surfaces liquidity hiding within existing network structure. We focus here on two applications of Cycles to deepening secondary market liquidity: first, as a compression layer between existing clearing participants and CCPs; and second, as a means to incorporate the liquidity of the trade credit network into formal settlement, extending market clearing beyond financial obligations and into real-economy financing.

\end{abstract}

\section{Introduction}

Current clearing models in secondary markets rely almost exclusively on central counterparties (CCPs) and cash or cash-like collateral. While this reliance mitigates counterparty credit risk, it has given rise to new, highly concentrated forms of systemic risk. Regulators and academics have extensively documented the ``too important to fail" nature of CCPs \cite{Wendt2015}, the procyclicality of their margining practices which can amplify market stress \cite{MaruyamaCerezetti2019, KingNesmithEtAl2023}, and the significant hidden costs imposed by clearing fragmentation \cite{benos2024cost}.

At the same time, secondary markets today leave vast pools of short-term liquidity in the form of accounts receivable and payable (AR/AP) outside formal settlement systems. The scale of this excluded liquidity is substantial: on the aggregate balance sheet of non-financial US businesses, accounts payable are three times larger than bank loans and fifteen times larger than commercial paper \cite{barrot2016trade,ivashina2018trade}. 

Here, we propose a new paradigm of clearing which simultaneously addresses both of these challenges, offering a path to both deepen the liquidity available for post-trade settlement and to broaden access to clearing services while mitigating risks. This paper makes the following contributions. 
\textbf{First}, we introduce a theoretical framework based on double-entry bookkeeping and graph theory to formally represent the economy as a network of interlocking balance sheets. This framework motivates a new clearing paradigm based on the atomic execution of multilateral settlement cycles across such a network of balance-sheets, which allows for a natural integration between trade-credit and financial markets. 
\textbf{Second}, we show mathematically how this cycle-finding approach compares to CCP-based systems. \textbf{Third}, we briefly sketch how the approach can be implemented in a privacy-preserving manner via distributed ledger technology as the Cycles Protocol \cite{buchman2025cycles}, using zero-knowledge proofs and Trusted Execution Environments. \textbf{Fourth}, we describe a stylized model of liquidity in secondary market clearing, using global multi-lateral setoff as an interoperability primitive across CCPs, financial intermediaries, and the trade credit network, synchronizing and reducing settlement liquidity needs across silos. \textbf{Fifth}, we examine empirical evidence from a real-world analogue, a simplified deployment of the clearing logic that demonstrates that linking specialized financial markets with a broad inter-firm credit network can substantially increase market liquidity.

We begin with a review of relevant background on both clearing systems and trade credit, and follow our contributions with a comparison of our solution to existing Liquidity Savings Mechanisms (LSMs), and a discussion of the implications of our framework and protocol for the economy and various stakeholders. Our contributions are at the intersection of new clearing and risk models and liquidity risk transmission. By re-framing trade credit as a potential liquidity source, the paper provides both academics and policymakers with a new perspective on how clearing networks can integrate real-economy obligations. We conclude by describing implications for supervisory frameworks, collateral management innovations, and the design of technology-enabled resilient settlement infrastructures.

\section{Background and Related Work}

Existing settlement infrastructures relevant here fall into two main domains: \emph{secondary market clearing}, which typically uses CCP novation and margining; and \emph{payment systems}, where RTGS operators and liquidity-saving mechanisms optimize settlement liquidity without CCP-style novation or loss mutualization. In secondary markets, the use of CCP novation improves market safety at the expense of high collateral costs, limited access for non-bank participants, and the introduction of new systemic risk vectors \cite{duffie2011does, ruffini2015central, KingNesmithEtAl2023}. 
Different CCPs specialize in different secondary market instruments, including derivatives, securities, and repo. All require collateral, novate trades, and generate settlement obligations with the CCP. In derivatives clearing, the collateral collected by CCPs comes in two forms: initial margin (IM), posted at trade inception and held until termination, and variation margin (VM), posted daily over the life of the contract to reflect mark-to-market (MTM) changes. VM is typically structured as final settlement of daily exposure (a settlement-transfer model, STM) \cite{ISDA2017VMSettlement}. Securities and repo market clearing may also generate mark-to-market settlement obligations, as well as final delivery.
The Cycles Protocol is a settlement-layer mechanism most applicable to mark-to-market settlement flows and to final delivery: it does not set margin, mutualize losses, or replace CCP default management. Its function is to reduce gross cash (or cash-like) movements needed to meet given settlement obligations via multilateral setoff over a broader obligation graph.

In payment systems, central banks have developed LSMs within RTGS to reduce intraday liquidity needs by identifying offsetting obligations prior to settlement. Although conceptually close to cycle-based optimization, LSMs are confined to payments in central bank reserves and cannot mobilize other assets. More generally, the value of reducing gross exposures through multilateral netting is well-established in the post-crisis financial literature. Often termed portfolio compression, this form of network optimization can dramatically reduce systemic obligations in OTC markets \cite{DErricoRoukny2021}, forming a key motivation for the development of more advanced, automated systems like the Cycles Protocol.

The Cycles Protocol generalizes the logic of LSMs while departing fundamentally from CCP clearing. It enables direct offsetting of existing obligations, without novation and without a CCP, so that trade credit (AR/AP) and financial market obligations can be settled in a single multilateral graph, deepening liquidity and broadening access. This positions Cycles as both a complement to existing CCP-based infrastructures and a novel design for mobilizing real-economy liquidity in secondary markets.

Trade credit represents the largest source of short-term financing. In total, accounts payable on the balance sheets of non-financial US firms are roughly three times larger than bank credit and fifteen times larger than commercial paper, illustrating the scale of inter-firm financing relative to formal finance \cite{ivashina2018trade,barrot2016trade}. Across advanced economies, trade credit balances are comparable to - or even exceed - short-term debt securities and loans, and typically account for around 20\% of GDP, a ratio that has remained remarkably stable over the past 25 years \cite{boissay2020trade,machokoto2022evolution}. This vast and underutilized reservoir of working capital finance forms the backbone of the liquidity of the real economy that resides outside formal clearing networks.

The macroeconomic evidence reinforces the stabilizing and amplifying role of trade credit in liquidity transmission. The presence of supplier-financed trade credit allows economies to sustain up to 14\% higher output relative to counterfactual spot economies and absorbs roughly 45\% of output losses during major financial shocks, such as those modeled to mimic the Great Recession. These effects arise because relational contracting enable suppliers to extend credit even when banks don't \cite{bocola2023macroeconomics}. As such, trade credit acts as distributed credit, an endogenous liquidity mechanism backed by reputation that complements formal financial credit channels, rather than competing with them. Historically, loss rates on trade receivables are significantly lower than in other asset classes, supporting the view that this decentralized credit network is resilient and information-rich \cite{boissay2020trade}.

Integrating trade credit into post-trade clearing through protocols such as Cycles would activate this dormant liquidity\footnote{In the same way that firms with more liquid equity provide a transmission channel for improved capital redistribution by converting their own improved access to financing into an extension of trade credit financing to others \cite{shang2020trade}, the inclusion of accounts receivable and payable within a multilateral clearing graph would transform working capital obligations into settlement resources.}.
This deepens market liquidity, may reduce settlement cash needs, can ease collateral pressures depending on institutional design, and extends participation to firms with productive but illiquid balance sheets. Empirically, since 80\% of trade credit exposures are domestic and 20\% cross-border, the mechanism naturally aligns with existing settlement jurisdictions \cite{boissay2020trade}. By leveraging this real-economy liquidity, post-trade systems can internalize a vast and stable source of short-term finance that would bridge the informational and institutional divide between production networks and financial markets.

Chile already provides a real-world example of the functioning of trade credit as financing through its Bolsa de Productos (BPC). 
According to the Chilean financial rule (CMF's General Regulation No. 429), small businesses can sell their outstanding invoices from approved creditworthy customers on a financial market. This applies both to individual invoices and to the official legal documents that represent the ownership of those invoices. This process allows the business to get paid immediately while ensuring that the debt is still legally valid and that the customer's payment goes to the new owner of the invoice. This mechanism effectively treats invoice obligations as securities that can be traded among investors, giving invoice owners cash-flow liquidity without relying on traditional bank credit or factoring. Academic and regulatory documentation confirms that the BPC market provides standardized clearing and notification procedures, insurance and guarantee of credit quality, and a legal framework (Law 19.220) that protects investors who acquire invoice instruments \cite{CMFChile2019InvoiceReg, boschmans2017fostering, carmona2010mercado}.

While the integration of trade credit into secondary markets is not a novel concept, its integration into secondary market \textit{clearing} is. That said, 
the underlying principle of multilateral trade credit settlement is not without a real-world precedent. A compelling and successful implementation has been operating at a national scale in Slovenia for decades \cite{schara2018deleveraging}. This system demonstrates that integrating specialized financial markets with the broad inter-firm credit network can dramatically deepen market liquidity and may improve liquidity coordination. This example suggests that the primary barriers to wider adoption in other jurisdictions have been less conceptual and more a result of institutional path dependency and the technological complexity of coordination, challenges which our protocol is designed to address.

The technological underpinning for such a multilateral, non-intermediated clearing system lies in recent advancements in Distributed Ledger Technology (DLT). However, our proposal is not a naive call for complete disintermediation. We share the cautious perspective of researchers who note the economic challenges and potential market concentration in settlement based on DLT \cite{BenosGarrattGurrolaPerez2017}, and who argue that the core risk management functions of a CCP cannot be easily replicated by current DLTs \cite{PlataChanCerezetti2024}. Therefore, the Cycles Protocol focuses on a specific and powerful function where DLT excels: atomic, multilateral settlement via balance sheet compression without novation. Furthermore, by incorporating privacy-preserving technologies (e.g. zero-knowledge proofs), it addresses the confidentiality concerns that have historically hindered the adoption of transparent on-chain systems for institutional and trade finance.

The Cycles approach extends clearing capabilities directly to non-bank firms in the trade credit network, without novation or membership at a CCP. Most non-bank firms today access CCPs indirectly through clearing members -- usually banks or broker-dealers -- that submit trades, interpose themselves operationally and legally between client and CCP, and provide liquidity services such as settlement, intraday credit, and margin financing. Liquidity stress in CCP-based markets is therefore transmitted to end-clients through intermediary balance sheets and operational capacity, not just through CCP rulebooks. In our context, any reduction in the cash need for gross settlement can mitigate this transmission channel even if the CCP risk management and membership is unchanged.

This transmission differs by market segment. In securities and repo clearing, the main immediate burden is funding gross settlement payments. In derivatives clearing, it is primarily VM cash flows. The IM is posted separately and lies outside our model. We therefore treat Cycles as a pre-novation settlement-efficiency layer that lowers gross settlement-cash needs for obligations routed through intermediaries, not as a substitute for CCP risk management, client clearing, or bank balance sheets.

\section{Theoretical Framework}

We base our clearing model on double entry accounting. Double-entry accounting allows for informal financing arrangements to become formalized and integrated into a clearing network. The key to this power is the use of the full functionality of accounting for the settlement of obligations.

Starting from the first principles of double entry accounting, one can observe that there are fundamentally four ways to discharge an outstanding obligation between two parties using accounting operations \cite{clavero2022money, buchman2025cycles}. The simplest and most well understood means is to pay the obligation via the assignment to the payee of the payer's title to an asset --- a transfer. Another commonly used method is to offset the outstanding obligation against another existing payment obligation of the payee to the payer. A third method is novation, where the outstanding obligation is discharged via the payer accepting another existing payment obligation of the payee. This is conditional on the payee's creditor agreeing to such a novation. The final method is issuance, where the outstanding payment obligation is discharged by issuance of a new obligation. The four ways to settle are depicted in the stylized balance sheets in tables \ref{tab:PayerPayee} and \ref{tab:FourWaysToSettle}.

%-----------------------
% Payer and Payee table
%-----------------------

\begin{table}[ht]
    \centering
    \begin{tabular}{l wc{1.5cm} | wc{1.5cm} c wc{1.5cm} | wc{1.5cm} c}
        %\toprule
        & \multicolumn{2}{c}{\textbf{Payer}} & & \multicolumn{2}{c}{\textbf{Payee}} & \\
        %\cmidrule{2-3} \cmidrule{5-6}
        & \small{Assets} & \small{Liabilities} & & \small{Assets} & \small{Liabilities} \\
        %\midrule
        \cmidrule{2-3} \cmidrule{5-6}
        $\Delta$ AR/AP &   & \EUR{-100} & & \EUR{-100} &   \\
        $\Delta$ funds  & A & B          & & C          & D \\
        %\bottomrule
    \end{tabular}
    \caption{Two balance sheets showing the transaction, with an additional column for delta notations.}
    \label{tab:PayerPayee}
\end{table}

Table \ref{tab:PayerPayee} shows two balance sheets with a desired result, a settled payment obligation of \EUR{100} in the bottom row. The desired change in Accounts Receivable and Payable ($\Delta$ AR/AP) must be matched with the change in funds ($\Delta$ funds) to maintain the balance of the balance sheets. Letters A to D indicate four possible positions in the balance sheets to create such a match.
Since every settlement involves two parties and each party has two accounts to use for settlement, we arrive at a two-by-two table \ref{tab:FourWaysToSettle} with four ways of settling the obligation.

%---------------------------
% Four ways to settle table
%---------------------------

\begin{table}[ht]
    \centering
    \renewcommand{\arraystretch}{1.5} % Adjust the spacing
    \begin{tabular}{cc}
                       & \textbf{Payee} \\
        \textbf{Payer} &
                        \begin{tabular}{ccc}
                            \toprule
                            Payment by & \cellcolor{gray!20} C=\EUR{+100} & \cellcolor{gray!20} D=\EUR{-100} \\
                            \midrule
                            \cellcolor{gray!20} A=\EUR{-100} & assignment & setoff \\
                            \cellcolor{gray!20} B=\EUR{+100} & issuance   & novation \\
                            \bottomrule
                        \end{tabular}
    \end{tabular}
    \caption{Four ways to settle.}
    \label{tab:FourWaysToSettle}
\end{table}

\subsection{From Balance Sheets to Network Representation}

In order to leverage this accounting insight to develop a novel clearing mechanism, we must progress from a bilateral to a multilateral view -- from a pair of balance sheets to a network of them. This section unfolds the concept of Settlement Flows by progressively building upon two key ideas: balance sheets and network representation. We will begin by representing the settlement mediums with their own separate balance sheets to provide a foundational understanding. Building upon this, we will then shift to visualizing the interactions between participants using a network of directed graphs. This network representation effectively captures the flow over the edges of both existing \textit{obligations}, which are claims already recorded on balance sheets from past activities, and \textit{acceptances}, which are agreed-upon commitments to potential new obligations. It is the integration of these forward-looking acceptances alongside historical debts that unlocks the full potential of double-entry bookkeeping for dynamic, multilateral clearing. This methodological approach, which maps accounting flows and stocks to graph-theoretic structures while preserving double-entry consistency, is a recognized technique in the analysis of complex financial systems \cite{fennell2014visualising}. Finally, we will examine how the network transforms after a settlement is complete, showing how past obligations are discharged while future-facing acceptances are converted into new obligations.

\begin{center}
\fbox{\parbox{0.9\textwidth}{%
    \textbf{Definition: Obligations and Acceptances}\\[0.5em]
    Edges in the network represent two types of claims:
    \begin{itemize}
        \item \textit{obligations} \textemdash{} enforceable debts that are already recorded on balance sheets, and
        \item \textit{acceptances} \textemdash{} conditional commitments to recognize new obligations upon settlement.
    \end{itemize}
}}
\end{center}

\subsubsection{Balance Sheet for Settlement Medium}

To present the settlement as a flow in full clarity, we must introduce a balance sheet for the settlement medium. A settlement medium is any asset or instrument used to discharge an obligation, including cash, bank deposits, stablecoins, tokenized assets, or other transferable claims. In the common case where the settlement medium consists of central bank reserves, the relevant balance sheet is that of the central bank. Without loss of generality, in our framework we represent any settlement medium as the liability of its issuer, thus incorporating the balance sheet of the issuer into the network.   
Table \ref{tab:DebtorCreditorSettlementMediumAssignment} shows a settlement by assignment, where the settlement medium is shown with a separate balance sheet. All flows required for such settlement are now flows between the balance sheets. This representation follows the representation logic developed by Nichols \cite{nichols1961modern} and published in a textbook by the Federal Reserve Bank of Chicago.

\begin{center}
\fbox{\parbox{0.9\textwidth}{\textbf{Definition: Settlement Flow} \\
A \emph{settlement flow} is a balanced sequence of debits and credits across balance sheets that extinguishes one or more obligations and transforms acceptances into new reciprocal obligations.}}
\end{center}

%--------------------------------------------------------------------------
% Debtor, Creditor, and Settlement Medium table with Settlement Flow
%--------------------------------------------------------------------------

\begin{table}[ht]
    \centering
    \begin{tabular}{l wc{1.1cm} | wc{1.1cm} c wc{1.1cm} | wc{1.1cm} c wc{1.1cm} | wc{1.1cm} c}
        & \multicolumn{2}{c}{\textbf{Settlement}} & & \multicolumn{2}{c}{} & & \multicolumn{2}{c}{} & \\
        & \multicolumn{2}{c}{\textbf{Medium} $S$} & & \multicolumn{2}{c}{\textbf{Debtor} $v_1$} & & \multicolumn{2}{c}{\textbf{Creditor} $v_2$} & \\
        & \small{Assets} & \small{Liabilities} & & \small{Assets} & \small{Liabilities} & & \small{Assets} & \small{Liabilities} \\
        \cmidrule{2-3} \cmidrule{5-6} \cmidrule{8-9}
        % First row
            \tikz[remember picture]\node (AC) {$\Delta$ AR/AP}; &
            & & & &
            \tikz[remember picture]\node (DAP) {\EUR{-100}}; &
            &
            \tikz[remember picture]\node (CAR) {\EUR{-100}}; &   \\
        % Second row
            \tikz[remember picture]\node (Funds) {$\Delta$ Debtor funds}; & 
            &
            \tikz[remember picture]\node (SDF) {\EUR{-100}}; & 
            &
            \tikz[remember picture]\node (DA) {\EUR{-100}}; & 
            & & & & \\
        % Third row
            \tikz[remember picture]\node (Funds) {$\Delta$ Creditor funds}; & 
            &
            \tikz[remember picture]\node (SCF) {\EUR{+100}}; & 
            & & & &
            \tikz[remember picture]\node (CA) {\EUR{+100}}; \\
    \end{tabular}
    \begin{tikzpicture}[overlay, remember picture, ->, >=stealth', shorten >=1pt, line width=1pt, auto]
        \draw[thick, draw=black, line width=2pt, bend left=33] (DAP) to (CAR);
        \draw[dotted, draw=blue, bend left=33] (CA) to (SCF);
        \draw[thick, draw=red, bend left=33] (SDF) to (DA);    
    \end{tikzpicture}
    \vspace{0.5cm}
    \caption{Balance sheet of Settlement Medium, Debtor, and Creditor showing the settlement by assignment, with an additional row for delta notations, and an indication of a cycle composed of tender, obligation, and acceptance.}
    \label{tab:DebtorCreditorSettlementMediumAssignment}
\end{table}

\subsubsection{Network Representation of Settlement Flow}

\begin{figure}[ht]
    \centering
    \begin{subfigure}[b]{0.45\textwidth}
        \centering
        \begin{tikzpicture}[->,>=stealth',shorten >=1pt,auto,node distance=2cm, thick,node/.style={circle,draw}]
            \node[node] (v1) {$v_1$};
            \node[node] (v2) [right of=v1] {$v_2$};
            \node[node] (v3) [below of=v1, xshift=1cm, yshift=0.4cm] {$S$};
            \path
                (v1) edge[line width=2pt, bend left=33] (v2)
                (v2) edge[dotted, line width=1pt, draw=blue, bend left=33] (v3)
                (v3) edge[line width=1pt, draw=red, bend left=33] (v1);
        \end{tikzpicture}
        \caption{Assignment}
        \label{fig:Assignment}
    \end{subfigure}
    \begin{subfigure}[b]{0.45\textwidth}
        \centering
        \begin{tikzpicture}[->,>=stealth',shorten >=1pt,auto,node distance=2cm, thick,node/.style={circle,draw}]
            \node[node] (v1) {$v_1$};
            \node[node] (v2) [right of=v1] {$v_2$};
            \node[node] (v3) [below of=v1, xshift=1cm, yshift=0.4cm] {$S$};
            \path
                (v1) edge[line width=2pt, bend left=33] (v2)
                (v2) edge[line width=1pt, draw=blue, bend left=33] (v3)
                (v3) edge[line width=1pt, draw=red, bend left=33] (v1);
        \end{tikzpicture}
        \caption{Setoff}
        \label{fig:setoff}
    \end{subfigure}
    \\
    \vspace{0.5cm}
    \begin{subfigure}[b]{0.45\textwidth}
        \centering
        \begin{tikzpicture}[->,>=stealth',shorten >=1pt,auto,node distance=2cm, thick,node/.style={circle,draw}]
            \node[node] (v1) {$v_1$};
            \node[node] (v2) [right of=v1] {$v_2$};
            \node[node] (v3) [below of=v1, xshift=1cm, yshift=0.4cm] {$S$};
            \path
                (v1) edge[thick, line width=2pt, bend left=33] (v2)
                (v2) edge[dotted, line width=1pt, draw=blue, bend left=33] (v3)
                (v3) edge[dotted, line width=1pt, draw=red, bend left=33] (v1);
        \end{tikzpicture}
        \caption{Issuance}
        \label{fig:Issuance}
    \end{subfigure}
    \begin{subfigure}[b]{0.45\textwidth}
        \centering
        \begin{tikzpicture}[->,>=stealth',shorten >=1pt,auto,node distance=2cm, thick,node/.style={circle,draw}]
            \node[node] (v1) {$v_1$};
            \node[node] (v2) [right of=v1] {$v_2$};
            \node[node] (v3) [below of=v1, xshift=1cm, yshift=0.4cm] {$S$};
            \path
                (v1) edge[thick, line width=2pt, bend left=33] (v2)
                (v2) edge[thick, line width=1pt, draw=blue, bend left=33] (v3)
                (v3) edge[dotted, line width=1pt, draw=red, bend left=33] (v1);
        \end{tikzpicture}
        \caption{Novation}
        \label{fig:Novation}
    \end{subfigure}
    \caption{Four settlement mechanisms as 3-node cycles in an obligation network.
                Nodes $v_1$ and $v_2$ are counterparties (payer and payee, respectively), and $S$ is the settlement medium (e.g., cash, a bank, a voucher/IOU issuer, or any balance-sheet entity). In each panel, settlement occurs by completing a directed cycle that can be executed atomically to discharge the original claim from $v_1$ to $v_2$.
                \textit{\small{Black solid arrow: existing payment obligation $v_1 \to v_2$. Red arrow: $v_1$'s offer/tender of settlement via $S$. Blue arrow: $v_2$'s consent/acceptance of $S$. Solid arrows denote existing obligations; dotted arrows denote acceptances which, upon execution, convert into obligations with reversed direction.}}}
    \label{fig:FourWaysToSettleNetworkFlow}
\end{figure}
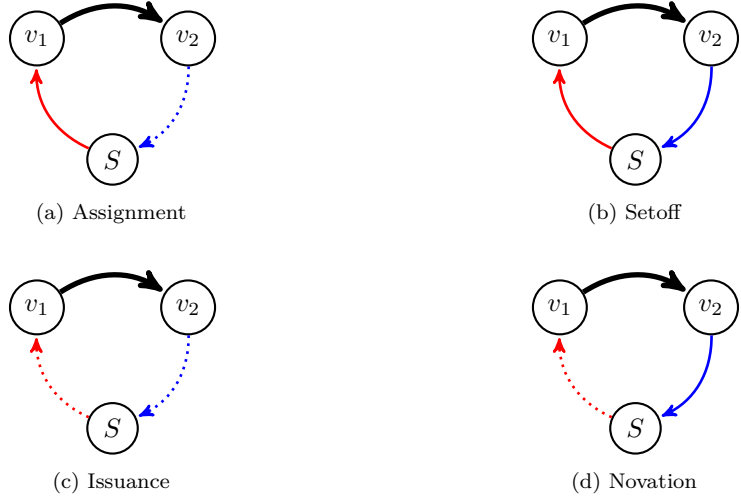

As a next step, we represent the balance sheets and flows in a simplified form. Balance sheets are the nodes, and flows are the edges of a directed graph. Arrows representing the edges can have a number indicating the amount in a chosen Settlement Unit of Account (SUoA). If all amounts on the graph are the same, we omit them to keep the graph simple. The beginning of the arrow shows where the amount is debited and the end where the amount is credited. The four ways to settle an obligation are presented in this way in Figure \ref{fig:FourWaysToSettleNetworkFlow}.

%---------------------------------------------------
% Four ways to settle with a Settlement Flow
%---------------------------------------------------

For the \textbf{Assignment} in figure \ref{fig:Assignment}, we have an existing payment obligation where $v_1$ must pay to $v_2$. The edge indicates that $v_1$ has a liability -- usually Accounts Payable (AP) -- and $v_2$ has an asset -- usually Accounts Receivable (AR). To make payment and settle the payment obligation, $v_1$ offers the settlement media $S$ he owns. The edge indicates that $S$ has a liability towards $v_1$. Better said, $S$ has an obligation towards $v_1$. For the Settlement Flow, we need an acceptance of $S$ as a settlement medium from $v_2$. This acceptance is a willingness of $v_2$ to accept $S$ as a settlement medium. This situation always occurs when we transfer money from one bank account to another. In a generalized case, $S$ can be any actor capable of holding assets and incurring liabilities.

In the case of the \textbf{setoff} in figure \ref{fig:setoff} the edge to the settlement medium $S$ is no longer a promise in the future as in the case of Assignment, but it is an already existing obligation. In the most simplified situation, one can think of $S$ as an existing obligation of $v_2$ toward $v_1$, represented as a balance sheet. In a more generalized scenario, $S$ can be another entity. Driving this further $S$ can represent a chain of obligations in a larger network of obligations.

For the \textbf{Issuance} in figure \ref{fig:Issuance} we can quickly recognize that $v_1$ offered $S$ from a loan to settle the payment obligation to $v_2$. If $v_2$ accepts $S$, the payment obligation is settled. Thinking outside of the banking scenario, $S$ can be a voucher issued by $v_1$. As long as such a voucher is accepted, it works as a settlement medium.

\textbf{Novation} in figure \ref{fig:Novation} looks counterintuitive at first glance. Instead of using an asset, the liability of the payee is used as a settlement medium. In case $v_1$ has nothing that $v_2$ is willing to accept, but on the contrary, $v_1$ is willing to accept the liability of $v_2$, the settlement flow can be created and the obligation of $v_1$ to pay $v_2$ can be settled. This is rarely used in daily practice by non-financial firms, since the person who controls $S$ must be happy with $v_1$ as a debtor instead of $v_2$. This requires additional coordination. However, the concept of a Settlement Network presents an opportunity to use this way to its full advantage without additional coordination.

What is obvious from the figures of all four settlement flows is that they are all \textit{cycles} -- a closed path of obligations and acceptances within the clearing network that can be executed atomically to discharge all included claims simultaneously.

\subsubsection{Network After Completion of Settlement}

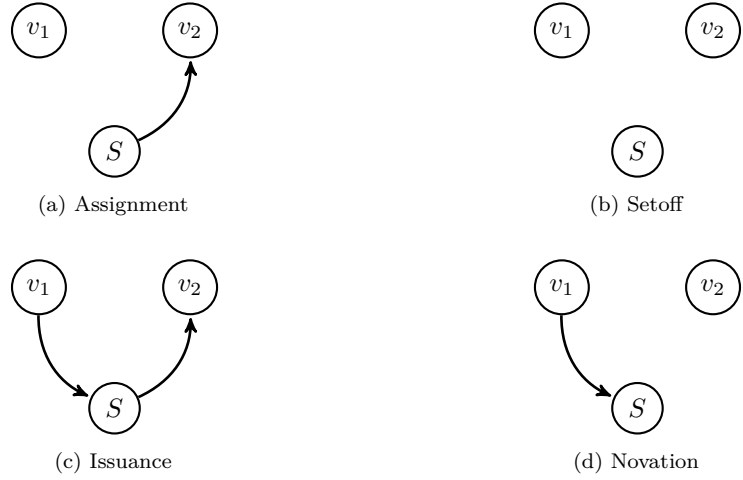
\begin{figure}[ht]
    \centering
    \begin{subfigure}[b]{0.45\textwidth}
        \centering
        \begin{tikzpicture}[->,>=stealth',shorten >=1pt,auto,node distance=2cm, thick,node/.style={circle,draw}]
            \node[node] (v1) {$v_1$};
            \node[node] (v2) [right of=v1] {$v_2$};
            \node[node] (v3) [below of=v1, xshift=1cm, yshift=0.4cm] {$S$};
            \path
                (v3) edge[thick, line width=1pt, bend right=33] (v2);
        \end{tikzpicture}
        \caption{Assignment}
        \label{fig:AssignmentAfter}
    \end{subfigure}
    \begin{subfigure}[b]{0.45\textwidth}
        \centering
        \begin{tikzpicture}[->,>=stealth',shorten >=1pt,auto,node distance=2cm, thick,node/.style={circle,draw}]
            \node[node] (v1) {$v_1$};
            \node[node] (v2) [right of=v1] {$v_2$};
            \node[node] (v3) [below of=v1, xshift=1cm, yshift=0.4cm] {$S$};
        \end{tikzpicture}
        \caption{Setoff}
        \label{fig:setoffAfter}
    \end{subfigure}
    \\
    \vspace{0.5cm}
    \begin{subfigure}[b]{0.45\textwidth}
        \centering
        \begin{tikzpicture}[->,>=stealth',shorten >=1pt,auto,node distance=2cm, thick,node/.style={circle,draw}]
            \node[node] (v1) {$v_1$};
            \node[node] (v2) [right of=v1] {$v_2$};
            \node[node] (v3) [below of=v1, xshift=1cm, yshift=0.4cm] {$S$};
            \path
                (v1) edge[thick, line width=1pt, bend right=33] (v3)
                (v3) edge[thick, line width=1pt, bend right=33] (v2);
        \end{tikzpicture}
        \caption{Issuance}
        \label{fig:IssuanceAfter}
    \end{subfigure}
    \begin{subfigure}[b]{0.45\textwidth}
        \centering
        \begin{tikzpicture}[->,>=stealth',shorten >=1pt,auto,node distance=2cm, thick,node/.style={circle,draw}]
            \node[node] (v1) {$v_1$};
            \node[node] (v2) [right of=v1] {$v_2$};
            \node[node] (v3) [below of=v1, xshift=1cm, yshift=0.4cm] {$S$};
            \path
                (v1) edge[thick, line width=1pt, bend right=33] (v3);
        \end{tikzpicture}
        \caption{Novation}
        \label{fig:NovationAfter}
    \end{subfigure}
    \caption{Obligation network after executing the settlement cycle in Figure~\ref{fig:FourWaysToSettleNetworkFlow}.
                Nodes $v_1$ and $v_2$ are counterparties and $S$ is the settlement medium. All arrows shown are \emph{resulting outstanding obligations} after completion. Acceptances have been executed and transformed into obligations with reversed direction.
                \textit{\small Takeaway: settlement discharges the original obligation $v_1 \to v_2$ and leaves (a) in \textbf{Assignment}, a new claim of $v_2$ on $S$; (b) in \textbf{Setoff}, no remaining obligations among $v_1,v_2,$ and $S$ (full cancellation in this minimal case); (c) in \textbf{Issuance}, a new claim of $v_2$ on $v_1$ via $S$ (i.e., $v_1$ has issued a liability accepted by $v_2$); and (d) in \textbf{Novation}, a new claim of $v_1$ on $S$ while $v_2$ now owes $S$ (the creditor is effectively substituted).}}
    \label{fig:FourWaysToSettleAfterSettlement}
\end{figure}

All four ways to settle look similar in the network representation. Settlement flows always involve cycles made up of existing obligations and acceptances. However, the result of a completed settlement flow has different effects on these elements. Obligations are fully or partially discharged, while acceptances transform into new obligations, but with the direction of the flow reversed. The four outcomes are depicted in Figure \ref{fig:FourWaysToSettleAfterSettlement}.

%----------------------------------
% Obligation network after Payment
%----------------------------------

Notice that the remaining full edges depicting obligations are in place of the previous dotted edges depicting acceptances, but in the opposite direction. In the case of assignment \ref{fig:AssignmentAfter}, the acceptance to store the value in the settlement medium $S$ offered by $v_1$ becomes a fact after the settlement. The entity represented by $v_2$ now has a deposit in $S$. In the case of issuance \ref{fig:IssuanceAfter}, we see that $v_1$ now owes $S$, and $S$ owes $v_2$. Thinking of $v_1$ and $v_2$ as firms and $S$ as a bank, it is easy to interpret the situation. The case of novation in \ref{fig:NovationAfter} also becomes easier to understand -- the initial liability of $v_2$ to $S$ is now a liability of $v_1$. But this holds in generalized situations, too. The nodes represent the balance sheets of various entities or persons.

\subsubsection{Obligations and Acceptances}

In the preceding sections, we distinguished between \textit{obligations} and \textit{acceptances} as the fundamental elements of settlement flows. An \textit{obligation} represents an enforceable claim, such as an account payable, while an \textit{acceptance} denotes a pre-commitment to receive or recognize a new obligation as part of settlement. Together, they form the dynamic structure through which value circulates in the network.

Conceptually, an acceptance can be viewed as a conditional obligation: once the agreed settlement flow is executed, it transforms into an enforceable liability. This distinction allows us to represent both existing debts and future commitments within the same graph, using solid edges for obligations and dotted edges for acceptances. When a settlement cycle is executed, all participating obligations are discharged, and acceptances are converted into new, reciprocal obligations—closing the loop of exchange.

This abstraction parallels the logic of contract execution, where an offer and an acceptance together form a binding agreement. In our context, the network formalism generalizes this bilateral logic to a multilateral setting, enabling atomic settlement across multiple parties and assets within a unified accounting graph.

\subsection{CCP Netting and Cycles Multilateral Setoff}

CCP clearing and the Cycles Protocol both tackle the same bottleneck, settling large networks of payment obligations, but use fundamentally different primitives to do so. A CCP achieves netting by novating trades in a hub-and-spoke structure, and embeds settlement inside a risk-management stack (initial margin, variation margin, default funds and CCP capital) \cite{veraart2025systemic, zhu2011there}. While the netting function can be performed without a CCP (as in payments clearing and CLS), novation with a CCP provides the most optimal netting outcome and allows the CCP to assume all counterparty credit risk. Cycles, by contrast, perform multilateral setoff directly on the obligation graph by canceling cycles, without novation, and without assuming counterparty credit risk. By finding and executing composite cycles on a network of balance sheets, it functions as a neutral settlement layer \cite{dini2025neutral}, discovering clearing paths that can cross the institutional divide between real-economy working capital and financial market transactions. Cycles can thus be used alongside CCPs to reduce the cash needed to meet \emph{payment obligations} (e.g., VM calls, final delivery), but it does not change the CCP's margin \emph{calibration} (IM/VM models) or replace margin as a credit-risk buffer \cite{ISDA2017VMSettlement}. We summarize as follows:

\begin{quote}\small
\textbf{Terminology (netting vs setoff).}
\begin{itemize}
    \item \textbf{Setoff} extinguishes obligations by offsetting claims against counterclaims. \textbf{Multilateral setoff} generalizes this to chains/cycles across multiple parties, discharging them simultaneously.
    \item \textbf{Netting} replaces multiple obligations with net payment obligations (who owes how much in total), but does not by itself discharge the underlying claims.
    \item \textbf{Compression} terminates economically redundant contracts to reduce gross notional and the number of outstanding obligations. Compression via multi-lateral set-off is \textit{conservative}, since it does not mutate counterparties, but only discharges cyclic flows. Compression via netting is \textit{nonconservative}, since net positions sum across counterparties \cite{DErricoRoukny2021}.
    \item \textbf{CCP multilateral netting} combines \emph{novation} with netting: bilateral trades are replaced by CCP-facing positions and then netted inside the CCP netting set, with counterparty credit risk managed via margining and the default waterfall.
\end{itemize}
\end{quote}

 In the rest of this section, we formalize the distinction between CCP netting and multilateral setoff. We start with a formal definition of a payment obligations network, netting, and multilateral setoff, then introduce a simple mutual default fund as a source of liquidity to compare netting and multilateral setoff.

\paragraph{Obligation network.}
Let $G=(V,E,p)$ be a directed weighted payment-obligation network, where $V$ is the set of participating firms with $n:=|V|$ and $E\subseteq V\times V$ is the set of bilateral obligations. Each edge $(u,v)\in E$ indicates that firm $u\in V$ owes firm $v\in V$ an amount $p_{uv}>0$. All payment obligations are strictly positive, $p:E\to\mathbb{R}_{++}$ with $p(u,v)=p_{uv}$, and no participant owes himself $u\neq v$. We index firms by $\mathcal{N}:=\{1,2,\ldots,n\}$.

\paragraph{Vector operations and net positions.}
For any vector $x\in\mathbb{R}^n$, define the positive part and $\ell_1$-norm by
\begin{align}
\label{eq1}
    \begin{split}
        x^+ &:= (\max\{x_1,0\},\max\{x_2,0\},\ldots,\max\{x_n,0\}), \\
        \|x\| &:= \sum_{i=1}^{n} |x_i|.
    \end{split}
\end{align}
Define the balance vector $b\in\mathbb{R}^n$ (net position of each firm) by
\begin{equation}
    b_i := \sum_{(u,i)\in E} p_{ui} - \sum_{(i,v)\in E} p_{iv}.
\end{equation}

\paragraph{CCP netting.}
Netting replaces bilateral obligations in $E$ by obligations to/from a CCP. CCP becomes a creditor to all net debtors and a debtor to all net creditors \cite{garvin2012central, anderson2014economics}. The total of payments required after netting and before drawing on any funds is
\begin{equation}
    N := \|b^+\|.
\end{equation}

In our stripped-down model for comparing netting with multilateral setoff, we assume that none of the participating firms holds liquid assets to meet their payment obligations. The CCP can draw on a mutual default fund of size $\delta\in\mathbb{R}_+$, which represents the total liquidity available to the network to discharge obligations. The remaining amount to be paid after the default fund is therefore a function of $\delta$,
\begin{equation}
    N(\delta) := \max\{0,\|b^+\|-\delta\}.
\end{equation}

\paragraph{Cycles multilateral setoff.}
Cycles discharges obligations in the payment-obligation network by computing a maximum circulation on an augmented graph that incorporates the default fund \cite{fleischman2021math}. Since the feasible set of discharged payment obligations and the optimum vary with $\delta$, we make this dependence explicit.

Let $s$ be a source-of-funds node and $t$ a sink. We define the $\delta$-augmented network as
\begin{equation}
    G^{\delta}:=(V^{\delta},E^{\delta},p^{\delta}),\qquad V^{\delta}:=V\cup\{s,t\},\qquad E^{\delta}:=E\cup E_{\delta}\cup\{(t,s)\},
\end{equation}
where $E_{\delta}$ denotes the set of edges linking participants to $s$ and $t$,
\begin{equation}
    E_{\delta}:=\{(s,i): b_i<0\}\cup\{(i,t): b_i>0\}.
\end{equation}
Net debtors have an edge from the source $s$ and net creditors have an edge to the sink $t$, each carrying their respective net positions,
\begin{equation}
    p^{\delta}_{si}:=-b_i,\qquad p^{\delta}_{it}:=b_i.
\end{equation}
Edges from net creditors to the sink $t$ are needed to complete circulation in the augmented obligation network, through an edge connecting the sink $t$ with the source $s$. The capacity of this $(t,s)$ edge defines the amount available in the default fund $\delta$,
\begin{equation}
    p^{\delta}_{ts}:=\delta.
\end{equation}
For all original obligation edges $(u,v)\in E$, capacities are unchanged: $p^{\delta}_{uv}:=p_{uv}$.

Let $c(\delta)=\{c_{uv}(\delta)\}_{(u,v)\in E^{\delta}}$ denote a maximum circulation on $G^{\delta}$, defined as the solution to
\begin{equation}
\label{eq:max-circulation}
\begin{aligned}
\text{maximize}\quad & \sum_{(u,v)\in E^{\delta}} c_{uv} \\
\text{subject to}\quad & \\
\text{capacity constraint}\quad & 0\le c_{uv}\le p^{\delta}_{uv}, && \forall (u,v)\in E^{\delta},\\
\text{and flow conservation}\quad & \sum_{v\in V^{\delta}} c_{uv} - \sum_{v\in V^{\delta}} c_{vu} = 0, && \forall u\in V^{\delta}.
\end{aligned}
\end{equation}

The discharge of the obligations by the circulation $c(\delta)$ leaves residual unpaid obligations on the original set of edges $E$. We define the total residual amount after Cycles multilateral setoff as
\begin{equation}
    S(\delta) := \sum_{(u,v)\in E}\bigl(p_{uv}-c_{uv}(\delta)\bigr).
\end{equation}

\begin{figure}[!htbp]
    \centering
    \captionsetup{font=footnotesize,skip=2pt} % local: smaller/tighter caption to help float placement
    \begin{subfigure}[c]{0.20\textwidth}
        \begin{subfigure}[b]{\textwidth}
            \centering
            \begin{tikzpicture}[->,>=stealth',shorten >=1pt,thick,scale=1.1,transform shape,
                node/.style={circle,draw,inner sep=1pt}]
                \def\r{1.1cm}
                \node[node] (v1) at (90:\r)   {$v_1$};
                \node[node] (v2) at (18:\r)   {$v_2$};
                \node[node] (v3) at (-54:\r)  {$v_3$};
                \node[node] (v4) at (-126:\r) {$v_4$};
                \node[node] (v5) at (162:\r)  {$v_5$};
                \path
                    (v1) edge[line width=1pt, draw=red, bend right=25] (v5)
                    (v5) edge[line width=1pt, draw=red, bend right=25] (v4)
                    (v4) edge[line width=1pt, draw=red, bend right=25] (v3)
                    (v3) edge[line width=1pt, draw=red, bend right=25] (v2)
                    (v2) edge[line width=1pt, draw=red, bend right=25] (v1)
                    (v3) edge[line width=0.9pt] (v1)
                    (v1) edge[line width=0.9pt] (v2)
                    (v3) edge[line width=0.9pt] (v5)
                    (v4) edge[line width=0.9pt] (v2)
                    (v5) edge[line width=0.9pt] (v1);
            \end{tikzpicture}
            \label{fig:Network}
        \end{subfigure}

        \par\smallskip
        \begin{subfigure}[b]{\textwidth}
            \centering
            \begin{tikzpicture}[->,>=stealth',shorten >=1pt,thick,scale=1.1,transform shape,
                node/.style={circle,draw,inner sep=1pt}]
                \def\r{1.1cm}
                \node[node] (v1) at (90:\r)   {$v_1$};
                \node[node] (v2) at (18:\r)   {$v_2$};
                \node[node] (v3) at (-54:\r)  {$v_3$};
                \node[node] (v4) at (-126:\r) {$v_4$};
                \node[node] (v5) at (162:\r)  {$v_5$};
                \path
                    (v1) edge[line width=1pt, draw=red, bend right=25] (v5)
                    (v5) edge[line width=1pt, draw=red, bend right=25] (v4)
                    (v4) edge[line width=1pt, draw=red, bend right=25] (v3)
                    (v3) edge[line width=1pt, draw=red, bend right=25] (v2)
                    (v2) edge[line width=1pt, draw=red, bend right=25] (v1);
            \end{tikzpicture}
            \label{fig:Cyclic}
        \end{subfigure}

        \par\smallskip
        \begin{subfigure}[b]{\textwidth}
            \centering
            \begin{tikzpicture}[->,>=stealth',shorten >=1pt,thick,scale=1.1,transform shape,
                node/.style={circle,draw,inner sep=1pt}]
                \def\r{1.1cm}
                \node[node] (v1) at (90:\r)   {$v_1$};
                \node[node] (v2) at (18:\r)   {$v_2$};
                \node[node] (v3) at (-54:\r)  {$v_3$};
                \node[node] (v4) at (-126:\r) {$v_4$};
                \node[node] (v5) at (162:\r)  {$v_5$};
                \path
                    (v3) edge[line width=0.9pt] (v1)
                    (v1) edge[line width=0.9pt] (v2)
                    (v3) edge[line width=0.9pt] (v5)
                    (v4) edge[line width=0.9pt] (v2)
                    (v5) edge[line width=0.9pt] (v1);
            \end{tikzpicture}
            \label{fig:Acyclic}
        \end{subfigure}
    \end{subfigure}%
    \hfill
    \begin{subfigure}[c]{0.78\textwidth}
        \centering
        \includegraphics[width=\textwidth,height=0.55\textheight,keepaspectratio]{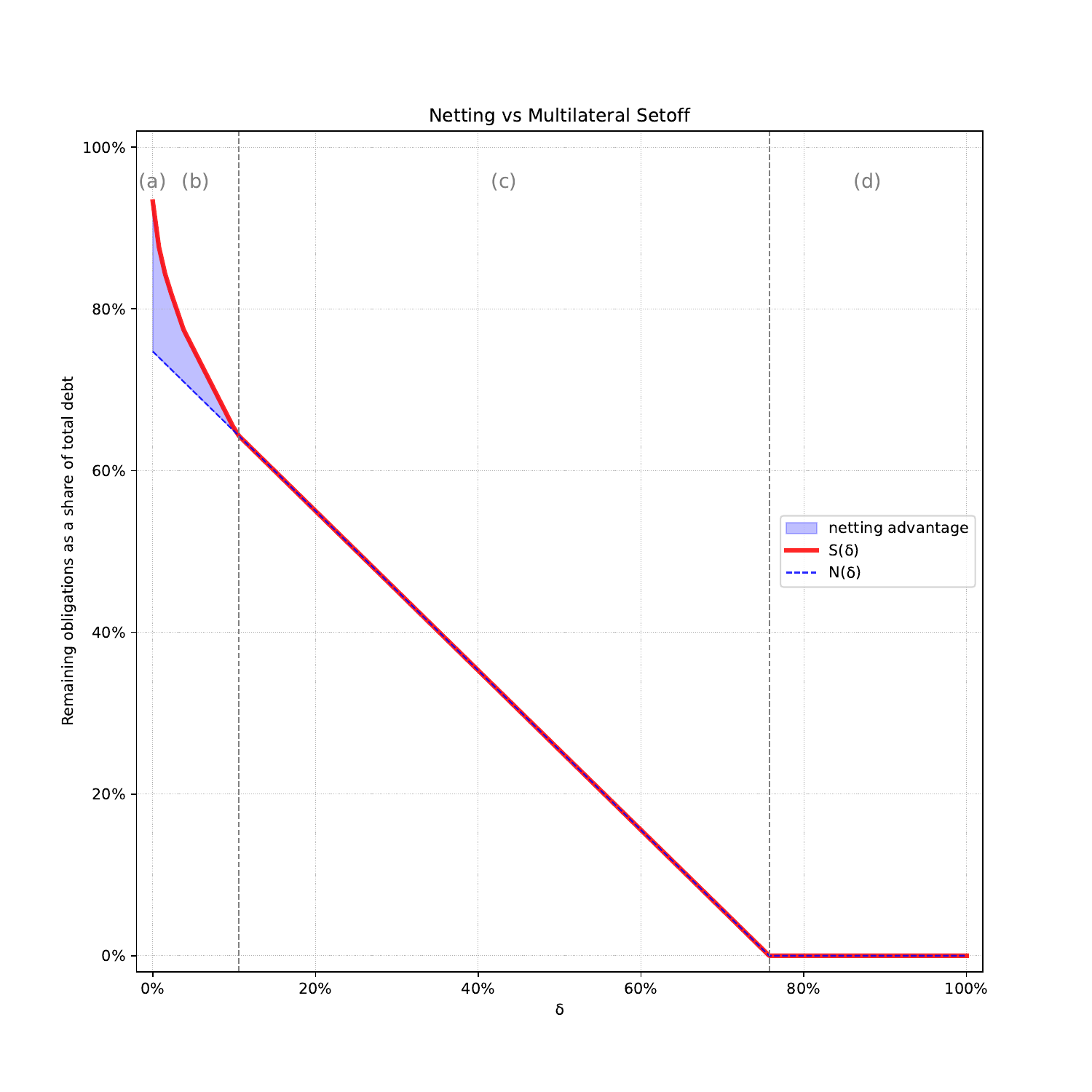}
    \end{subfigure}
    \caption[Comparing CCP Netting with Cycles Multilateral Setoff]{Comparing CCP netting with Cycles multilateral setoff as available liquidity increases.
        LHS: stylized obligation network among five participants $v_1,\dots,v_5$. The red directed cycle shows obligations discharged by multilateral setoff; the black edges are the remaining acyclic obligations that require additional liquidity to be discharged.
        RHS: total residual obligations after settlement under CCP netting, $N(\delta)$, and Cycles multilateral setoff, $S(\delta)$, as a function of exogenous liquidity $\delta$.
        \textit{\small{Takeaway: CCP netting may dominate at very low $\delta$, but Cycles quickly closes the gap as $\delta$ rises by routing liquidity through long obligation chains.}} \\ \scriptsize Source: AJPES (data available from the authors upon request).}
    \label{fig:NettingVsCycles}
\end{figure}
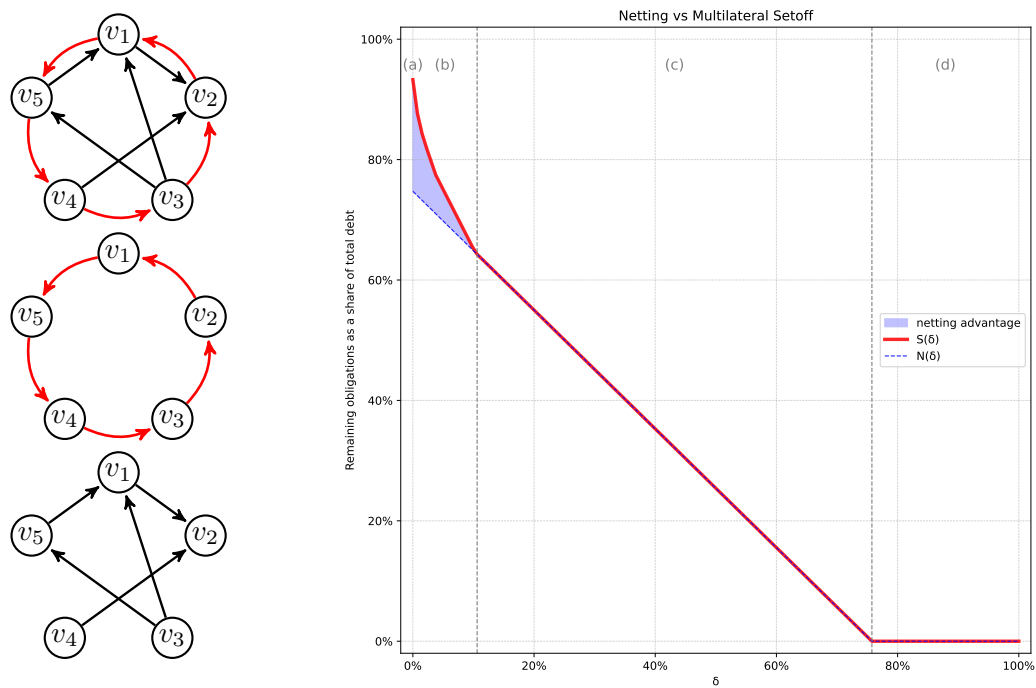

\paragraph{Comparing Netting and Multilateral Setoff.} Figure \ref{fig:NettingVsCycles} shows $N(\delta)$ and $S(\delta)$ as $\delta$ varies in this simplified model. It provides a stylized analytical comparison of netting and multilateral setoff under simplified assumptions. The left side shows a five‑participant network, with the cyclic part in red discharged by multilateral offset and the remaining acyclic part in black. The right side shows the total residual amounts after the CCP netting, $N(\delta)$, and the multilateral setoff of the Cycles, $S(\delta)$. The results are based on a real trade credit network among thousands of Slovenian businesses, chosen for ease of comparison. A network of payment obligations after exchange trading would have the same shape, but both curves would shift down and left toward the origin, since an exchange-trading network clears more than a trade credit network. 

We can observe that both $N(\delta)$ and $S(\delta)$ follow the same path most of the time. There is a distinct advantage for CCP netting when there is no liquidity available from the default fund ($\delta=0$), but this advantage vanishes quickly as the available funding increases.

Comparison of netting and multilateral setoff reveals four operating modes for the use of funds in different operating regimes depicted by vertical dashed lines and labeled (a) to (d) in figure \ref{fig:NettingVsCycles}.

At starting point (a) no funds are available. Netting clearly outperforms multilateral setoff but concentrates risk in the CCP. With Cycles multilateral setoff, risk stays with trading partners, so the cleared amount is much smaller than under CCP netting. Figures \ref{fig:CCP_a} and \ref{fig:Acyclic_a} show the obligation network under both CCP netting and Cycles multilateral setoff.

As funds become available, Cycles improves quickly because its objective is to maximize circulation. The available funds are first pushed through the longest paths in the remaining payment obligation network. Figures \ref{fig:CCP_b} and \ref{fig:Acyclic_b} show the network in this operating mode. This second mode terminates when $S(\delta) = N(\delta)$, where Cycles setoff matches CCP netting, the network has no more chains, and there are no further network-based gains in discharged amounts. What remains are disconnected individual payment obligations. 
In the mode shown in figures \ref{fig:CCP_c} and \ref{fig:Acyclic_c}, the $45^o$ lines represent straightforward use of funds to settle these individual obligations, with no network effects. Finally, in mode (d), available funds exceed funding needs, so all obligations are fully paid and nothing remains outstanding, as shown in figures \ref{fig:CCP_d} and \ref{fig:Acyclic_d}.

\begin{figure}[ht]
    \centering
    \begin{subfigure}[b]{0.24\textwidth}
        \centering
        \begin{tikzpicture}[->,>=stealth',shorten >=1pt,thick,scale=1.1,transform shape,
            node/.style={circle,draw,inner sep=1pt}]
            \def\r{1.1cm}
            \node[node] (v1) at (90:\r)   {$v_1$};
            \node[node] (v2) at (18:\r)   {$v_2$};
            \node[node] (v3) at (-54:\r)  {$v_3$};
            \node[node] (v4) at (-126:\r) {$v_4$};
            \node[node] (v5) at (162:\r)  {$v_5$};
            \node[diamond,draw,inner sep=0.1pt,minimum size=2.2mm,font=\tiny] (c) at (0,0) {$CCP$};
            \path
                (c) edge[line width=0.9pt] (v1)
                (c) edge[line width=0.9pt] (v2)
                (v3) edge[line width=0.9pt] (c)
                (v4) edge[line width=0.9pt] (c)
                (v5) edge[line width=0.9pt] (c);
        \end{tikzpicture}
        \caption{}
        \label{fig:CCP_a}
    \end{subfigure}
    \begin{subfigure}[b]{0.24\textwidth}
        \centering
        \begin{tikzpicture}[->,>=stealth',shorten >=1pt,thick,scale=1.1,transform shape,
            node/.style={circle,draw,inner sep=1pt}]
            \def\r{1.1cm}
            \node[node] (v1) at (90:\r)   {$v_1$};
            \node[node] (v2) at (18:\r)   {$v_2$};
            \node[node] (v3) at (-54:\r)  {$v_3$};
            \node[node] (v4) at (-126:\r) {$v_4$};
            \node[node] (v5) at (162:\r)  {$v_5$};
            \node[diamond,draw,inner sep=0.1pt,minimum size=2.2mm,font=\tiny] (c) at (0,0) {$CCP$};
            \path
                (c) edge[line width=0.9pt] (v1)
                (c) edge[line width=0.9pt] (v2)
                (v3) edge[line width=0.9pt] (c)
                (v5) edge[line width=0.9pt] (c);
        \end{tikzpicture}
        \caption{}
        \label{fig:CCP_b}
    \end{subfigure}
    \begin{subfigure}[b]{0.24\textwidth}
        \centering
        \begin{tikzpicture}[->,>=stealth',shorten >=1pt,thick,scale=1.1,transform shape,
            node/.style={circle,draw,inner sep=1pt}]
            \def\r{1.1cm}
            \node[node] (v1) at (90:\r)   {$v_1$};
            \node[node] (v2) at (18:\r)   {$v_2$};
            \node[node] (v3) at (-54:\r)  {$v_3$};
            \node[node] (v4) at (-126:\r) {$v_4$};
            \node[node] (v5) at (162:\r)  {$v_5$};
            \node[diamond,draw,inner sep=0.1pt,minimum size=2.2mm,font=\tiny] (c) at (0,0) {$CCP$};
            \path
                (c) edge[line width=0.9pt] (v1)
                (v3) edge[line width=0.9pt] (c);
        \end{tikzpicture}
        \caption{}
        \label{fig:CCP_c}
    \end{subfigure}
    \begin{subfigure}[b]{0.24\textwidth}
        \centering
        \begin{tikzpicture}[->,>=stealth',shorten >=1pt,thick,scale=1.1,transform shape,
            node/.style={circle,draw,inner sep=1pt}]
            \def\r{1.1cm}
            \node[node] (v1) at (90:\r)   {$v_1$};
            \node[node] (v2) at (18:\r)   {$v_2$};
            \node[node] (v3) at (-54:\r)  {$v_3$};
            \node[node] (v4) at (-126:\r) {$v_4$};
            \node[node] (v5) at (162:\r)  {$v_5$};
            \node[diamond,draw,inner sep=0.1pt,minimum size=2.2mm,font=\tiny] (c) at (0,0) {$CCP$};
        \end{tikzpicture}
        \caption{}
        \label{fig:CCP_d}
    \end{subfigure}
    \caption{CCP netting “waterfall” across four funding regimes (panels a–d) in a five-member market. As prefunded liquidity support $\delta$ rises, obligations contract from many member-to-CCP pay/receive positions (a) to fewer obligations (b–c), and ultimately to full settlement with no remaining obligations in this stylized example (d).}
    \label{fig:CCP_watterfall}
\end{figure}
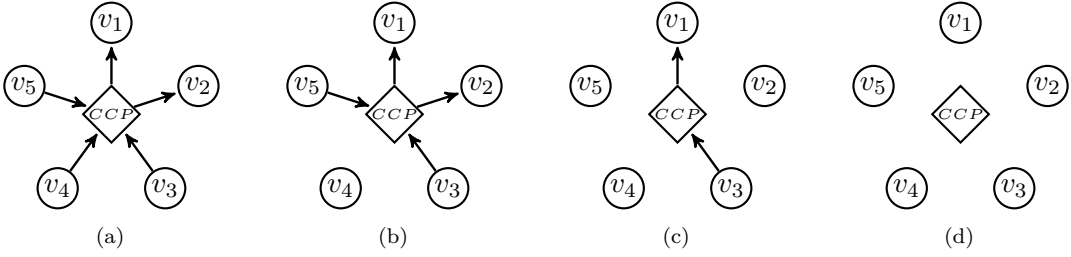

\begin{figure}[ht]
    \centering
    \begin{subfigure}[b]{0.24\textwidth}
        \centering
        \begin{tikzpicture}[->,>=stealth',shorten >=1pt,thick,scale=1.1,transform shape,
            node/.style={circle,draw,inner sep=1pt}]
            \def\r{1.1cm}
            \node[node] (v1) at (90:\r)   {$v_1$};
            \node[node] (v2) at (18:\r)   {$v_2$};
            \node[node] (v3) at (-54:\r)  {$v_3$};
            \node[node] (v4) at (-126:\r) {$v_4$};
            \node[node] (v5) at (162:\r)  {$v_5$};
            \path
                (v3) edge[line width=0.9pt] (v1)
                (v1) edge[line width=0.9pt] (v2)
                (v3) edge[line width=0.9pt] (v5)
                (v4) edge[line width=0.9pt] (v2)
                (v5) edge[line width=0.9pt] (v1);
        \end{tikzpicture}
        \caption{}
        \label{fig:Acyclic_a}
    \end{subfigure}
    \begin{subfigure}[b]{0.24\textwidth}
        \centering
        \begin{tikzpicture}[->,>=stealth',shorten >=1pt,thick,scale=1.1,transform shape,
            node/.style={circle,draw,inner sep=1pt}]
            \def\r{1.1cm}
            \node[node] (v1) at (90:\r)   {$v_1$};
            \node[node] (v2) at (18:\r)   {$v_2$};
            \node[node] (v3) at (-54:\r)  {$v_3$};
            \node[node] (v4) at (-126:\r) {$v_4$};
            \node[node] (v5) at (162:\r)  {$v_5$};
            \path
                (v3) edge[line width=0.9pt] (v1)
                (v1) edge[line width=0.9pt] (v2)
                (v3) edge[line width=0.9pt] (v5)
                (v5) edge[line width=0.9pt] (v1);
        \end{tikzpicture}
        \caption{}
        \label{fig:Acyclic_b}
    \end{subfigure}
    \begin{subfigure}[b]{0.24\textwidth}
        \centering
        \begin{tikzpicture}[->,>=stealth',shorten >=1pt,thick,scale=1.1,transform shape,
            node/.style={circle,draw,inner sep=1pt}]
            \def\r{1.1cm}
            \node[node] (v1) at (90:\r)   {$v_1$};
            \node[node] (v2) at (18:\r)   {$v_2$};
            \node[node] (v3) at (-54:\r)  {$v_3$};
            \node[node] (v4) at (-126:\r) {$v_4$};
            \node[node] (v5) at (162:\r)  {$v_5$};
            \path
                (v3) edge[line width=0.9pt] (v1);
        \end{tikzpicture}
        \caption{}
        \label{fig:Acyclic_c}
    \end{subfigure}
    \begin{subfigure}[b]{0.24\textwidth}
        \centering
        \begin{tikzpicture}[->,>=stealth',shorten >=1pt,thick,scale=1.1,transform shape,
            node/.style={circle,draw,inner sep=1pt}]
            \def\r{1.1cm}
            \node[node] (v1) at (90:\r)   {$v_1$};
            \node[node] (v2) at (18:\r)   {$v_2$};
            \node[node] (v3) at (-54:\r)  {$v_3$};
            \node[node] (v4) at (-126:\r) {$v_4$};
            \node[node] (v5) at (162:\r)  {$v_5$};
        \end{tikzpicture}
        \caption{}
        \label{fig:Acyclic_d}
    \end{subfigure}
    \caption{Cycles multilateral setoff “waterfall” across four funding regimes (panels a–d) in the same five-participant example as Figure~\ref{fig:CCP_watterfall}. Unlike CCP netting, Cycles first eliminates all cycles via multilateral setoff, leaving an acyclic residual network to be settled with external liquidity. As available liquidity/funding $\delta$ rises, these acyclic obligations are progressively discharged (a$\to$d). The early benefit derives from \emph{cycle removal} (balance-sheet compression) rather than risk concentration in a central counterparty.}
    \label{fig:Cycles_watterfall}
\end{figure}
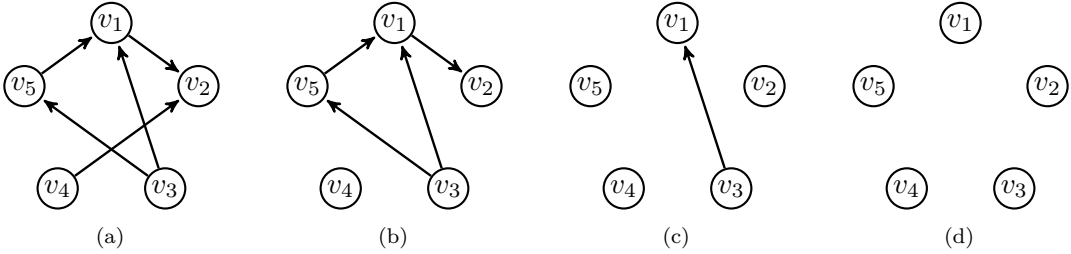

With this comparison of netting and setoff in mind, we can now assess their roles in risk management and propose a combined operating model. CCPs use novation to standardize and concentrate risk for netting and default management, while the Cycles Protocol uses multilateral setoff to compress balance sheets without assuming counterparty credit risk. Institutional separation between CCPs creates silos that can cause netting failures and concentrate operational and liquidity stress\cite{KingNesmithEtAl2023}. Joint clearing membership can also transmit liquidity shocks, for example through fire-sales of illiquid collateral or variation margin gains haircutting \cite{zhu2011there}. In our setting, Cycles provides a neutral settlement layer across silos by extinguishing interlocking obligations atomically \cite{dini2025neutral}. We therefore propose running Cycles’ multilateral setoff before CCP netting to shrink gross obligations before novation into fragmented CCP structures. \emph{Analytically,} this sequencing lowers the gross settlement liquidity needed for a given set of payment obligations.

The scope of clearing can therefore be increased beyond the management of risk through collateral posted to CCPs by incorporating the potential for systemic risk reduction through compression of interconnected past and future balance sheet claims.
Because Cycles operates without novation, it does not assume counterparty credit risk or provide CCP-style loss mutualization and default management. These remain as critical functions provided by CCPs. Cycles serves as a complement to those functions in markets where CCPs exist, and as an accessible form of liquidity-saving in markets where they do not (trade-credit).

\section{The Cycles Protocol Design}

The Cycles Protocol 
architecture is fundamentally different from the hub-and-spoke model of a CCP, which relies on novation to standardize and concentrate risk. Instead, the protocol establishes a peer-to-peer network where participants can express their financial relationships as a directed graph of balance sheet claims. The core function of the protocol is to discover and execute multilateral settlement flows within this graph, allowing for direct compression of the balance sheet among participants. This design choice inherently avoids the creation of a new systemic single point of failure and shifts the focus from managing concentrated risk to reducing overall obligations.

The power of the Cycles Protocol framework originates from its two primary primitives: obligations and acceptances.
By representing both past debts and future commitments on the same graph, the protocol creates a unified clearing space. This allows for the construction of settlement paths that can seamlessly cross institutional and economic boundaries, connecting the liquidity dormant in real-economy supply chains directly with the settlement needs of secondary financial markets.

Settlement is achieved through the discovery and atomic execution of closed loops or ``cycles" within the network graph. A specialized cycle detection algorithm \cite{fleischman2021math} continuously searches the graph for opportunities where a set of obligations and acceptances form a multilateral solvent clearing path. Once a valid cycle is identified, its settlement is executed as a single atomic transaction. Atomicity, a core feature enabled by the protocol's underlying distributed ledger technology, guaranties that all legs of the cycle settle simultaneously or none settle at all. This eliminates principal risk and the need for complex, sequential settlement instructions that characterize traditional correspondent banking, enabling true delivery-versus-payment (DvP) across multiple assets and participants in one indivisible operation.

Two additional design features are critical to the protocol's function as a practical market infrastructure: asset neutrality and privacy. The protocol is agnostic to the underlying settlement medium; while obligations and acceptances are denominated in a single unit of account, settlement flows can include any verifiable digital asset, including tokenized commercial bank money, stablecoins, or central bank digital currencies. This asset neutrality allows the integration of fragmented settlement media into a single, interoperable clearing layer. Furthermore, the protocol is designed to be privacy-preserving. Using zero-knowledge cryptography and trusted execution environments, participants can express their obligations and acceptances to the network without revealing confidential details of their transactions to the public. This ensures that while the network can mathematically verify the integrity of a settlement cycle, the underlying commercial relationships remain confidential, addressing a key prerequisite for institutional adoption.

The cumulative result of these design choices is a fundamentally new mechanism for post-trade clearing.
The Cycles Protocol complements the CCP's process of novation, risk concentration, and collateralization with a decentralized process of discovery, atomic execution, and direct balance sheet compression. Unlike CCPs that transfer risk through novation, Cycles extinguishes obligations through atomic multilateral setoff. It does not novate contracts, does not interpose itself as counterparty, and does not provide margining or default management—functions that remain with CCPs and intermediaries.
By providing an open, neutral, and privacy-preserving layer for multilateral settlement, the protocol is designed to unlock dormant pools of real-economy liquidity, reduce dependence on scarce high-quality collateral, and create a more resilient and accessible financial market infrastructure. More details on the protocol design and privacy-preserving architecture can be found in \cite{buchman2025cycles}. To summarize the core principles:

\begin{quote}\small
\begin{itemize}
    \item \textbf{Multilateral setoff and Compression}. The Cycles Protocol consolidates obligations and acceptances into a single obligation graph, enabling atomic multilateral settlement cycles that directly offset interconnected balance-sheet claims, maximizing balance-sheet compression and reducing gross settlement flows and liquidity needs.

    \item \textbf{Absence of Novation and Guaranty}. Cycles operates without novation or a CCP. In traditional systems, a CCP becomes the buyer to every seller and the seller to every buyer, guaranteeing contracts through novation. Cycles instead extinguishes obligations via atomic multilateral setoff, rather than transferring risk through novation.

    \item \textbf{Focus on Settlement Liquidity}. Cycles reduces settlement liquidity needs by using offsetting cycles. It generalizes LSMs from RTGS systems, extending liquidity optimization from interbank payments to diverse obligations like accounts receivable and payable.
\end{itemize}
\end{quote}

\section{Stylized Model: Closing the Liquidity Cycle}

Fragmentation in post-trade liquidity is, at its core, a symptom of incomplete collateral mobilization. When multiple CCPs clear similar products, market participants are forced to post redundant collateral, trapping liquidity and increasing costs \cite{benos2024cost}. Even as market execution spreads across venues, settlement liquidity remains narrowly defined in cash or cash-like assets.
Our approach addresses this issue directly by creating a unified clearing graph where obligations from different venues and even different economic sectors (trade credit) can be cleared against each other, unlocking the liquidity trapped by institutional and asset-class silos.

To translate the theoretical framework developed in the preceding sections into a concrete application, we now construct a stylized model 
that covers only settlement flows and variation margin-like payment legs. Initial margin and other CCP risk-management resources, such as default waterfalls, are out of scope.

This model will proceed in two stages to progressively demonstrate the power of the Cycles Protocol to overcome market fragmentation and unlock dormant liquidity. We begin by examining a common scenario involving two distinct CCPs, showing how the protocol generates significant liquidity savings even in a purely financial context. We then expand this model to incorporate a Trade Credit Network (TCN), illustrating the protocol's core contribution: the ability to mobilize real-economy balance sheets to deepen and stabilize secondary market settlement.

\subsection{Scope and limits of the model}
The stylized model in this section is intentionally narrow and focuses on \emph{settlement liquidity}. Analytically, it shows that multilateral setoff can reduce the \emph{gross cash} required to discharge a fixed set of payment obligations by canceling closed chains and routing available liquidity through remaining paths. Any implications for collateral usage, liquidity buffers, or margin levels are not modeled and depend on the specific CCP rulebook and margin methodology. Likewise, effects on initial margin calibration, procyclicality of margin calls, default probabilities, contagion across CCPs, and system-wide resilience are hypotheses that require pilot testing and empirical validation in realistic operational settings.

\subsection{Separate CCPs}

\begin{figure}[ht]
    \centering
    \begin{tikzpicture}[->,>=stealth',shorten >=1pt,auto,node distance=2cm, thick,node/.style={circle,draw}]
            \node[node] (A) {$A$};
            \node[node] (M) [above of=A] {$ \$ $};
            \node[node] (B) [above of=M] {$B$};
            \node[diamond,draw] (CCP1) [left of=M] {$CCP1$};
            \node[diamond,draw] (CCP2) [right of=M] {$CCP2$};
            \path
                (CCP1) edge[thick, line width=1pt, bend left=33] (B)
                (B)    edge[thick, line width=1pt, bend left=33] (CCP2)
                (CCP2) edge[thick, line width=1pt, bend left=33] (A)
                (A)    edge[thick, line width=1pt, bend left=33] (CCP1)
                (M)    edge[thick, line width=1pt, bend left=33] (A)
                (M)    edge[thick, line width=1pt, bend left=33] (B)
                (A)    edge[dotted, line width=1pt, bend left=33] (M)
                (B)    edge[dotted, line width=1pt, bend left=33] (M);
        \end{tikzpicture}
    \caption{A minimal settlement cycle with two non-interoperable CCPs (CCP1, CCP2) and two traders ($A$, $B$), shown as a two-node slice of a larger clearing network. Solid arrows show existing payment obligations from cleared trades (e.g., $A\to CCP1$, $CCP1\to B$, $B\to CCP2$, $CCP2\to A$), forming a closed loop across silos. The \$ node is the settlement asset (cash or equivalent). Dotted arrows ($A\to\$$, $B\to\$$) show traders’ acceptance of the asset. \textit{\small{Takeaway: with fragmented clearing, obligations are settled separately in each CCP, requiring gross liquidity. Cycles can settle cross-silo cycles atomically, extinguishing linked obligations and cutting external liquidity needs to at most the net imbalance.}}}
    \label{fig:CyclesCCP}
\end{figure}
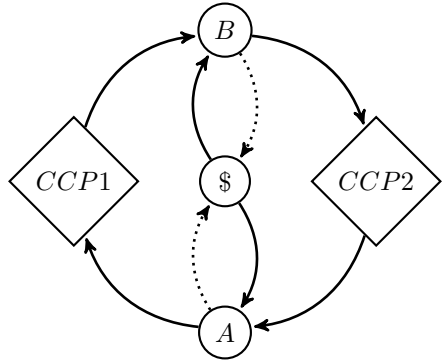

To illustrate the practical implications of our framework for the clearing of the secondary market, we present a stylized model depicted in Figure \ref{fig:CyclesCCP}. This model represents a common yet highly inefficient scenario in today's financial markets, which are characterized by clearing fragmentation. We consider two market participants, Trader A and Trader B, who have conducted trades across two separate and non-interoperable Central Counterparties, CCP1 and CCP2. Their resulting settlement obligations form a closed loop: Trader A owes a payment to CCP1, who in turn owes a payment to Trader B; simultaneously, Trader B owes a payment to CCP2, who in turn owes a payment to Trader A. This creates a chain of interlocking payables and receivables that are trapped across institutional silos.

\begin{figure}[ht]
    \centering
    % markets_ccps figure
    \includegraphics[width=\textwidth]{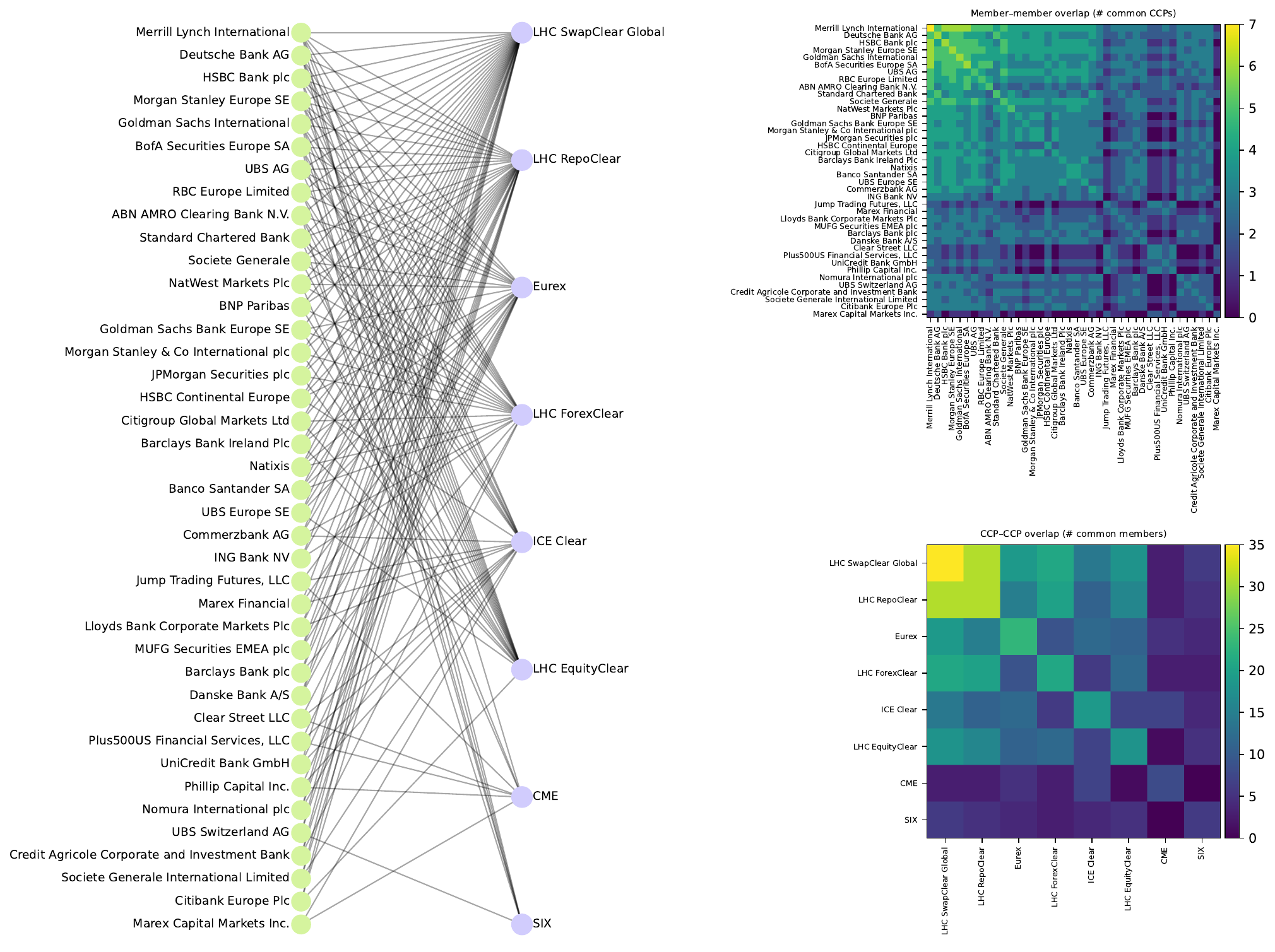}
    \caption{Member - CCP network. Heatmaps depict the number of CCPs members share (top right) and number of members CCPs share (bottom right). Figure is based on Veraart, L. A. M. (2025). ``Systemic risk in central counterparty networks". [Open access] Licensed under CC BY 4.0. \cite{veraart2025systemic}. The figure displays only the densest part of the CCP–members network and is illustrative rather than complete. \\ \scriptsize{Source: Authors collected publicly available member lists from CCPs’ websites in February 2026.}}
    \label{fig:Members_CCPs}
\end{figure}

In realistic markets with many clearing members and CCPs, membership overlaps are only partial and settlement obligations form a large, interconnected graph with multiple cycles. CCP netting works only within each CCP's own netting set, so cross-CCP cycles remain when obligations alternate between silos. Cycles addresses these residual cross-silo cycles by applying multilateral setoff to the integrated obligation graph. Figure~\ref{fig:Members_CCPs} shows that such partial overlap and multi-CCP connectivity are common in practice.

\begin{quote}\small
\noindent\textbf{Proposition: incremental benefit arises only across netting-set partitions.} In an obligation graph whose edges are split into disjoint netting sets (e.g., obligations cleared in CCP1 versus CCP2), any directed cycle that alternates between these sets cannot be fully eliminated by silo-by-silo CCP netting, whereas multilateral setoff on the integrated graph can discharge the entire cycle.

\noindent\textbf{Proof sketch.} CCP netting aggregates obligations \emph{within} each partition into net positions against that CCP. Because the cycle alternates across partitions, each partition sees only a path segment, not a closed loop, so those gross obligations remain as CCP-facing payables/receivables after silo netting. In the integrated graph, the same obligations form a closed cycle with zero net at every node. Multilateral setoff cancels all gross legs simultaneously, leaving only any residual net imbalance to fund.
\end{quote}

Under the current, fragmented clearing infrastructure, these obligations must be settled independently within each CCP's domain. Trader A must obtain liquidity to meet its gross payment obligation to CCP1, even though it holds an offset receivable from CCP2. This institutional separation prevents the use of one's assets to discharge one's liabilities across non-interoperable silos, creating a \textit{netting failure} at the level of the overall obligation graph. The consequence is a significant and unnecessary demand for scarce settlement assets and collateral.\footnote{Benos et al. \cite{benos2024cost} document costs of CCP fragmentation primarily through initial margin; our stylized example isolates the separate (and complementary) channel of fragmented \emph{settlement} cash flows. D'Errico \& Roukny \cite{DErricoRoukny2021} study compression in OTC networks in the context of the ``netting failures" that result from fragmented clearing services; they show that compression can reduce gross obligations and overcome this effect.}
Liquidity is trapped because the system lacks a holistic view of the underlying obligation network.

The Cycles Protocol provides a more efficient solution by providing a neutral settlement layer that can operate across these institutional silos. As shown in the figure, the protocol is able to view the entire graph of obligations and acceptances. The solid lines represent existing obligations (e.g., A $\rightarrow$ CCP1), while the dotted lines (A $\rightarrow$ \$, B $\rightarrow$ \$) represent the acceptance of the settlement asset by the participants. The core function of the protocol is to identify the multilateral settlement cycle (A $\rightarrow$ CCP1 $\rightarrow$ B $\rightarrow$ CCP2 $\rightarrow$ A). By executing this cycle as a single atomic transaction, all four interlocking obligations are simultaneously discharged through multilateral setoff.

The impact on systemic liquidity can be substantial. Let us assume that each obligation in the cycle is for \$100. In the fragmented model, Trader A and Trader B would collectively need to provide \$200 in cash to settle their respective gross payables. In the integrated model enabled by the Cycles Protocol, the entire chain of obligations is extinguished without any movement of the underlying settlement asset, reducing the net liquidity requirement. If the trades were for different amounts (e.g. A owed \$100 to CCP1 but was owed \$90 from CCP2), only the net difference of \$10 would need to be settled using external liquidity. This mechanism of balance sheet compression can mitigate settlement-liquidity stress associated with gross cash movements \cite{KingNesmithEtAl2023}.
Importantly, while Cycles reduces settlement liquidity needs, it does not affect the underlying counterparty credit risk of the original trades. The credit risk management function remains with the CCPs and their margining regimes.
Our model's conclusion is consistent with the broader literature on compression in financial networks \cite{DErricoRoukny2021}, who demonstrate that compression can reduce gross notional obligations.

Figure \ref{fig:Members_CCPs} shows how the simple stylized setup in Figure \ref{fig:CyclesCCP} becomes more complex in a broader network of CCPs and clearing members, where cycles involving three or more CCPs can easily arise. These cycles offer an additional, complementary tool for managing interdependencies, alongside members’ risk-management tools and CCP interconnection mechanisms \cite{Wendt2015}.

\subsection{Trade Credit Network as a liquidity source}

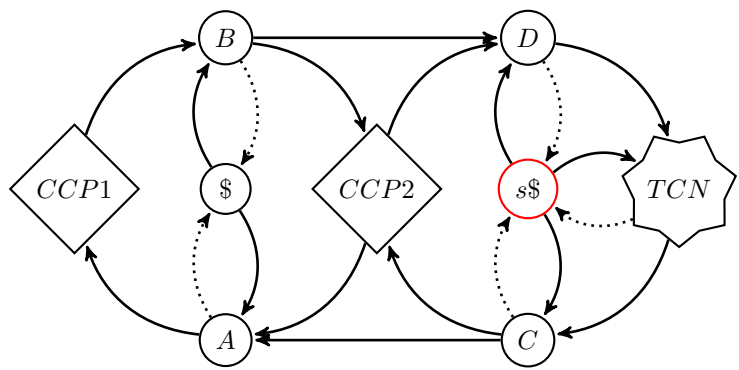
\begin{figure}[ht]
    \centering
    \begin{tikzpicture}[->,>=stealth',shorten >=1pt,auto,node distance=2cm, thick,node/.style={circle,draw}]
            \node[node] (A) {$A$};
            \node[node] (M) [above of=A] {$ \$ $};
            \node[node] (B) [above of=M] {$B$};
            \node[diamond,draw] (CCP1) [left of=M] {$CCP1$};
            \node[diamond,draw] (CCP2) [right of=M] {$CCP2$};
            \node[node] (SC) [right of=CCP2, draw=red] {$s\$ $};
            \node[node] (D) [above of=SC] {$D$};
            \node[node] (C) [below of=SC] {$C$};
            \node[star,star points=7,star point ratio=0.8,draw] (TCN) [right of=SC] {$TCN$};
            \path
                (CCP1) edge[thick, line width=1pt, bend left=33] (B)
                (B)    edge[thick, line width=1pt, bend left=33] (CCP2)
                (CCP2) edge[thick, line width=1pt, bend left=33] (A)
                (A)    edge[thick, line width=1pt, bend left=33] (CCP1)
                (M)    edge[thick, line width=1pt, bend left=33] (A)
                (M)    edge[thick, line width=1pt, bend left=33] (B)
                (A)    edge[dotted, line width=1pt, bend left=33] (M)
                (B)    edge[dotted, line width=1pt, bend left=33] (M)
                (B)    edge[thick, line width=1pt] (D)
                (C)    edge[thick, line width=1pt] (A)
                (CCP2) edge[thick, line width=1pt, bend left=33] (D)
                (C)    edge[thick, line width=1pt, bend left=33] (CCP2)
                (D)    edge[thick, line width=1pt, bend left=33] (TCN)
                (TCN)  edge[thick, line width=1pt, bend left=33] (C)
                (SC)   edge[thick, line width=1pt, bend left=33] (D)
                (SC)   edge[thick, line width=1pt, bend left=33] (C)
                (SC)   edge[thick, line width=1pt, bend left=33] (TCN)
                (D)    edge[dotted, line width=1pt, bend left=33] (SC)
                (C)    edge[dotted, line width=1pt, bend left=33] (SC)
                (TCN)  edge[dotted, line width=1pt, bend left=33] (SC)
                ;
        \end{tikzpicture}
    \caption{Integrating CCP settlement silos with a Trade Credit Network (TCN) to mobilize real-economy liquidity. Nodes $A$ and $B$ are traders; $CCP1$ and $CCP2$ are central counterparties. Nodes $C$ and $D$ are CCP-facing participants – banks or broker-dealers that intermediate between $TCN$ (star) clients and CCPs. The \$ circle is the traditional cash settlement asset for CCP obligations; the red node $s\$$ is a separate settlement asset in the trade-credit domain (e.g., stablecoin or tokenized cash). Solid arrows are existing obligations (payables); dotted arrows are acceptances/willingness to deliver the settlement asset. \textit{\small{Takeaway: representing CCP and trade-credit obligations in one unified clearing graph lets the Cycles Protocol detect and atomically execute longer cross-domain cycles (e.g., $A\to CCP1\to B\to D\to TCN\to C\to A$), so trade-credit receivables/payables offset settlement needs and reduce external cash liquidity.}}}
    \label{fig:CyclesCCPandTCN}
\end{figure}

The true potential of our new clearing paradigm is realized when we broaden the scope beyond financial markets to include real-economy obligations. Figure \ref{fig:CyclesCCPandTCN} depicts this expanded ecosystem. Here, we introduce two CCP-facing intermediaries, C and D (e.g., banks or broker-dealers acting as clearing members), who connect CCP settlement obligations to a broader Trade Credit Network (TCN). This network represents the vast web of accounts receivable and payable that constitute the primary source of working capital finance for non-financial firms. Furthermore, this commercial ecosystem can utilize its own settlement media, such as a tokenized asset or a stablecoin ($s\$$), creating yet another liquidity silo separate from the traditional ($\$$) cash settlement used by CCPs.

It is critical to clarify that the TCN is not a formal, secondary marketplace for invoices, but rather a digital representation of the \textit{inter-firm credit graph} formed by existing, organic commercial relationships. To integrate this heterogeneous network into a high-assurance settlement system, the protocol addresses the inherent challenges of credit quality and legal enforceability through a two-pronged approach. First, all participants operate under a unified legal framework or \textbf{cover contract}, which pre-authorizes the multilateral setoff and ensures the legal finality and enforceability of any settlement cycle executed by the protocol. Second, the protocol itself enforces a \textbf{double ascertainment} mechanism for every trade credit obligation. Before an account receivable/payable can be included in the clearing graph, its existence, amount, and key terms must be digitally confirmed and signed by both the creditor and the debtor. This process eliminates unilateral errors, prevents the inclusion of disputed or fraudulent invoices, and transforms each informal credit relationship into a verified, mutually-attested, and settlement-ready obligation.

In this interconnected but fragmented environment, there are new obligations. For example, Trader A has a commercial receivable from firm C (C $\rightarrow$ A), and Trader B has a payable to firm D (B $\rightarrow$ D). 
For simplicity, C and D denote CCP-facing participants -- banks or broker-dealers that intermediate between TCN clients and CCPs.
In the conventional model, these relationships are completely separate from their CCP settlement activities. The receivable C $\rightarrow$ A is an illiquid balance sheet item for Trader A. It cannot be used to satisfy its immediate cash requirement to pay CCP1. Similarly, the TCN's internal obligations (D $\rightarrow$ TCN and TCN $\rightarrow$ C) represent a self-contained pool of liquidity that is inaccessible to the broader financial market. This institutional and technological separation is precisely what our framework is designed to overcome.

The Cycles Protocol, acting as a neutral settlement layer, can map all these disparate obligations onto a single, unified clearing graph. By doing so, it can discover far more complex and efficient settlement paths that traverse these previously impermeable boundaries. For example, a longer cycle now becomes visible: A $\rightarrow$ CCP1 $\rightarrow$ B $\rightarrow$ D $\rightarrow$ TCN $\rightarrow$ C $\rightarrow$ A. This cycle atomically links a financial market obligation (A $\rightarrow$ CCP1), an inter-firm commercial debt (B $\rightarrow$ D), obligations internal to the TCN, and a commercial debt owed back to a financial market trader (C $\rightarrow$ A). The execution of this single, multilateral transaction nets out every link in the chain simultaneously.

The implication is that the Trade Credit Network now acts as a source of liquidity for the financial settlement system. Trader A's need to source external cash (\$) to pay CCP1 can potentially be completely eliminated: its payment obligation can be discharged, at least in part, by routing it through the network and offsetting it against the commercial receivable owed by C. This benefit is in the same spirit as Duffie and Zhu's recommendation to reduce counterparty exposures by jointly clearing standardized interest-rate swaps and credit default swaps within a single clearing house \cite{duffie2011does}. In our setting, the Cycles Protocol provides an interoperability layer that approximates this joint clearing across otherwise separate trading venues, CCPs, and commercial credit networks. As more venues and CCPs connect, they become liquidity sources for one another, enabling multilateral setoff that reduces net exposures across all participating infrastructures. The latent, illiquid value locked in the TCN's balance sheets is thus mobilized to resolve settlement bottlenecks in the secondary market, deepening the pool of settlement assets and broadening participation in a more resilient, efficient and interconnected ecosystem.

\section{Empirical Evidence from an Integrated Trade Credit Market}

The stylized models presented in the previous sections are not merely a conceptual exercise. A real-world precedent that demonstrates the power of integrating specialized financing with a broad trade credit network has been successfully operating in Slovenia for three decades. This system, described in detail by Fleischman \& Dini \cite{fleischman2020logic}, involves specialized Trade Credit Finance Markets (TCFMs) that act as crucial liquidity providers to the national Trade Credit Network (TCN). TCFMs provide funding to TCN debtors and offer accounts receivable collection services to creditors, effectively bridging liquidity gaps within the real-economy credit graph. The matching of TCFMs deals done over TCN offers a compelling empirical case for our proposal.

The evidence in this section is descriptive and drawn from a real-world analogue rather than a direct deployment of the Cycles Protocol. It shows that large-scale multilateral clearing over inter-firm trade credit networks is feasible, and that linking such a network to specialized liquidity providers can materially increase discharged obligations. It does \emph{not} show the performance, security, or governance properties of any specific protocol implementation nor identify causal effects or welfare impacts. The observed results reflect the design of the Slovenian system, participant selection, credit policies, and operating constraints.

\subsection{Dataset}

The analysis uses complete administrative clearing records from the Slovenian Agency for Public Legal Records and Related Services (AJPES), covering 111 monthly clearing cycles from May 2016 to March 2025. Table~\ref{tab:summary_stats} summarises the main characteristics of the full dataset and the snapshot month used for the network visualisation.

\begin{table}[ht]
\centering
\caption{Summary statistics of the Slovenian TCN clearing dataset}
\label{tab:summary_stats}
\begin{tabular}{lrr}
\hline
\textbf{Metric} & \textbf{Full dataset (2016--2025)} & \textbf{Snapshot (May 2016)} \\
\hline
Clearing cycles          & 111        & 1 \\
Unique firms             & 80{,}124   & 23{,}895 \\
Total obligations        & 5{,}880{,}447 & 94{,}683 \\
Avg.\ monthly obligations & 52{,}977  & --- \\
Avg.\ monthly volume     & €269M      & €381M \\
Avg.\ obligation size    & €5{,}464   & €4{,}020 \\
Avg.\ obligation size (2016) & €4{,}217 & --- \\
Avg.\ obligation size (2025) & €7{,}320 & --- \\
LP liquidity share\textsuperscript{a}       & 8.9\% $\pm$ 0.8\%  & 8.3\% \\
LP share of cleared value\textsuperscript{b} & 27.6\% $\pm$ 2.7\% & 27.6\% \\
Avg.\ network path length\textsuperscript{c} & 4.26               & 4.36 \\
Adjusted LP contribution\textsuperscript{d} & 37.9\% $\pm$ 3.3\% & 36.1\% \\
\hline
\multicolumn{3}{l}{\scriptsize \textsuperscript{a} Share of total credit volume submitted by identified TCFMs (rounded-amounts proxy).} \\
\multicolumn{3}{l}{\scriptsize \textsuperscript{b} Directly measured share of total cleared (setoff) value attributable to TCFM obligations.} \\
\multicolumn{3}{l}{\scriptsize \textsuperscript{c} Average clearing path length, $\ell = \ln N / \ln\ln N$, for a scale-free network \cite{barabasi1999emergence}.} \\
\multicolumn{3}{l}{\scriptsize \textsuperscript{d} LP liquidity share $\times$ average path length; accounts for the fact that each unit of LP} \\
\multicolumn{3}{l}{\scriptsize \quad liquidity clears $\ell$ edges as it propagates through the network.} \\
\multicolumn{3}{l}{\scriptsize Source: AJPES administrative clearing records (data available from the authors upon request).} \\
\end{tabular}
\end{table}

The dataset shows that the network shrank markedly during the observation period, from about 24,000 firms in 2016 to about 11,000 in 2025, while the average obligation size nearly doubled from €4,217 to €7,320. This suggests a maturation of the market. Smaller firms exited, and the remaining firms handled larger transactions. Despite fewer participants, the average monthly clearing volume remained between €200M and €380M, indicating sustained economic activity in the network.

The contribution of TCFM to the total cleared value is quantified as follows. TCFMs are identified via the rounded-amount proxy (obligations divisible by €10{,}000), and their direct credit volume is approximately 8.9\% of total monthly obligations ($\pm$0.8\%). This is only the direct share, while the \emph{effective} contribution is much larger. In a scale-free network \cite{barabasi1999emergence}, each unit of LP liquidity propagates through clearing cycles, clearing obligations along every edge it traverses. For a scale-free network with $N$~nodes, the average path length is $\ell = \ln N / \ln\ln N$ \cite{barabasi1999emergence}. With an average of $N \approx 16{,}000$ firms per cycle, $\ell \approx 4.26$. Multiplying the direct LP share by this path length yields an adjusted LP contribution of about \textbf{37.9\%} (std 3.3\%, range 31.8\%--47.2\%) across all 111~months, matching field estimates of 40--50\% and indicating that this is a structural market feature rather than an artifact of the snapshot month.

\begin{figure}[ht]
    \centering
    \includegraphics[width=1\linewidth]{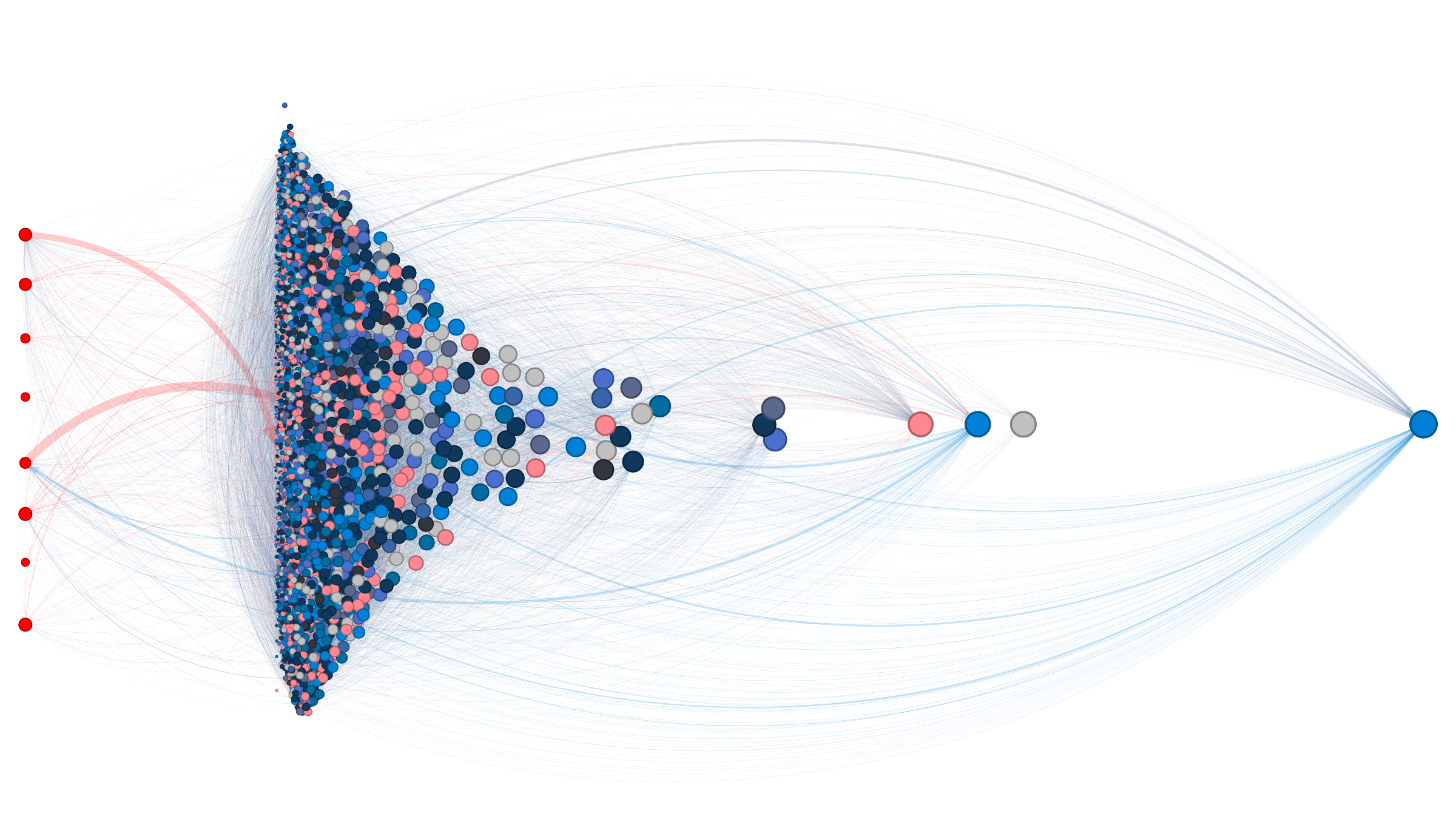}
    \caption[A visualization of the integrated Trade Credit Finance Markets (TCFMs) and Trade Credit Network (TCN) in Slovenia]{Integrated Trade Credit Finance Markets (TCFMs) and the Trade Credit Network (TCN) in Slovenia, May 2016 monthly clearing network. Red nodes: TCFMs (liquidity providers/market-makers); blue: TCN participants (mainly SMEs); salmon: firms active in both layers (bridges between TCFMs and the wider TCN). Node size reflects the number of obligations cleared that month. \textit{\small{Takeaway: specialized liquidity providers and a small set of bridge firms link into the dense SME trade-credit graph, enabling extensive multilateral setoff that would be impossible in a fragmented network.}} \\ \scriptsize Source: AJPES (data available from the authors upon request).}
    \label{fig:TCNclear}
\end{figure}

\subsection{Network structure}

Figure~\ref{fig:TCNclear} provides a visualization of the ecosystem for the snapshot month. The network graph shows the TCFMs as red nodes on the far left, acting as entry points for liquidity. The dense cluster in the center is made up of thousands of TCN participants (shades of blue), primarily SMEs. The salmon-colored nodes represent firms that are active in both the TCN and the TCFMs, acting as bridges between the two domains. The size of each node is proportional to the number of obligations it has cleared. This visual representation highlights the structure of the market, distinguishing between the numerous smaller participants and the larger, more central players.

The structure of the network reveals the depth of the market. On the right side of the graph, a few large firms play an outsized role, acting as anchors for a significant volume of clearing cycles. However, the greatest impact comes from the integration of the TCFMs on the left. Without the liquidity injection and market-making services provided by the TCFMs -- and the stability provided by the large anchor firms -- the dense but fragmented network of SMEs would not provide sufficient clearing paths.

\subsection{Worst-case scenario: liquidity provider exit}

To quantify the structural importance of TCFMs beyond the contribution estimate, we simulate the counterfactual in which LP firms exit the network entirely. For each of the 111~clearing cycles we identify LP firms as those that appear on \emph{both} sides of LP obligations --- as debtor in at least one LP transaction and as creditor in at least one --- a criterion that isolates true market-making intermediaries from firms that merely happen to submit a single round-amount obligation. We then remove all obligations involving any identified LP firm and re-run the MTCS clearing algorithm on the reduced network.

LP firms represent approximately 12.8\% of total submitted debt (std 2.7\%), yet their exit causes a disproportionate collapse in clearing. Across all 111~months, total cleared value falls by an average of \textbf{55.5\%} (std 5.8\%, range 31.8\%--71.2\%, median 54.7\%). In the snapshot month, clearing falls from €25.6M to €13.4M -- a drop of 47.7\%. The asymmetry between the debt share removed (12.8\%) and the clearing lost (55.5\%) reflects the structural role of LP firms as bridges within the network: their obligations participate in long clearing cycles that span many other firms, so their removal severs not just their own transactions, but the cycles those transactions enable.

This counterfactual provides a lower-bound estimate of LP importance: because the rounded-amounts proxy is conservative (some LP transactions may not use exactly round denominations), the true number of LP firms and the true clearing loss may be larger. However, even the conservative estimate confirms that LP firms are not peripheral contributors but load-bearing nodes whose presence is necessary for the network to function near its clearing potential.

\subsection{Interpretation}

The combination of Table~\ref{tab:summary_stats}, Figures~\ref{fig:TCNclear} -- \ref{fig:NettingVsCycles} and the counterfactual LP-exit constitutes an illustrative empirical analogue for the central mechanism of the Cycles Protocol. Several findings stand out. First, large-scale multilateral setoff over inter-firm trade credit networks is demonstrably feasible: the Slovenian system has operated continuously for nearly three decades, processing tens of thousands of obligations per clearing cycle across tens of thousands of firms. Second, the integration of specialized liquidity providers with the broader network is a structurally stable feature: accounting for the scale-free path-length multiplier ($\ell \approx 4.26$, \cite{barabasi1999emergence}), the adjusted TCFM contribution to cleared value averages 37.9\% (range 31.8\%--47.2\%) across all 111~months, consistent with field observations of 40--50\%. Third, LP firms are load-bearing nodes: their complete exit causes clearing to collapse by an average of 55.5\% despite representing only 12.8\% of submitted debt, confirming that their value lies in enabling long clearing cycles rather than in the volume of their own obligations. 
This descriptive evidence is consistent with our central thesis that in this setting, tighter integration between TCFMs and the TCN coincides with higher activity in both layers, with TCFMs facilitating more closed financing deals and the TCN achieving greater debt discharge.

Although the Slovenian system demonstrates that integrated trade-credit clearing can operate at scale, it was built with the technology of its time and relies on batch-based processes. Recent advances in distributed ledgers, smart contracts, and privacy-preserving computation motivate testing whether similar clearing logic can be implemented in a real-time setting with stronger automation and auditability. We treat this as a hypothesis for pilot deployments rather than a result established by the evidence in this section.

\section{Comparison with Liquidity-Saving Methods (RTGS)}

The dramatic liquidity efficiencies demonstrated in our stylized model invite a natural comparison with LSMs in RTGS systems. The literature on LSMs has long established their ability to reduce the intraday liquidity and collateral needs of the system \cite{JurgilasMartin2010_LSM_RTGS, GalbiatiSoramaki2010_LSM_BankBehaviour}. However, a deeper analysis reveals that their most significant contribution is not merely the mechanical efficiency of netting, but their impact on the strategic behavior of participants.

The primary benefit of an LSM is that it acts as a coordination device \cite{atalay2008welfare}. By providing an incentive for banks to submit payment obligations earlier, an LSM helps to resolve the system-wide gridlock that arises from strategic payment delays. The Cycles Protocol can be understood as a powerful generalization of this core principle. While an LSM addresses the coordination failure of \textit{payment timing} within a homogenous, single-asset environment (central bank reserves), the Cycles Protocol is designed to solve a far broader and more complex coordination failure: the inability to settle \textit{heterogeneous obligations} (trade credit, securities, and payments) across institutionally and technologically siloed networks.

The Cycles Protocol differs from traditional LSMs in its fundamental architecture. While an LSM optimizes flows within a single liquidity silo (central bank money), Cycles is designed as a neutral layer to bridge multiple heterogeneous liquidity silos. The key distinctions are threefold. First, its scope extends beyond interbank payments to include composite obligations such as trade credit and DvP settlement. Second, its mechanism is not a queuing overlay but a native atomic setoff, achieving finality through direct balance sheet compression rather than deferred gross settlement. Third, it allows settlement in a richer set of assets, including tokenized commercial bank money and stablecoins, not just central bank reserves.

Therefore, the Cycles Protocol can be seen as a powerful generalization of the LSM principle. It preserves the efficiency gains of multilateral offsetting but extends them beyond the banking system into the real economy and secondary markets. Where LSMs reduce liquidity needs for banks, Cycles unlocks dormant liquidity from commercial balance sheets, reduces collateral-locking across the entire economy, and helps address market fragmentation by deepening the pool of eligible settlement resources.

Like LSMs, Cycles optimizes settlement liquidity. It does not assume or reallocate counterparty credit risk, which remains with bilateral relationships and any CCP guarantees. Instead, Cycles targets inefficiencies outside CCPs’ core credit risk role. Fragmented clearing across multiple CCPs is costly because dealers cannot net positions across them, increasing liquidity demand. Cycles addresses this by creating a unified clearing graph for cross-asset, cross-institution obligations.
Although this promises significant welfare gains, it also introduces new design and regulatory challenges, including the cross-jurisdictional enforceability of obligations and the management of credit risk in a decentralized environment.

Table \ref{tab:lsm_comparison} summarizes the conceptual differences between LSMs and the Cycles Protocol. While both aim to optimize liquidity, Cycles generalizes the mechanism from centralized queuing to decentralized atomic setoff.

\begin{table}[ht]
    \centering
    \scriptsize
    \renewcommand{\arraystretch}{1.2}
    \setlength{\tabcolsep}{5pt}
    \begin{tabularx}{\textwidth}{|>{\raggedright\arraybackslash}p{2cm}|X|X|}
        \hline
        \rowcolor{gray!10}
        \textbf{Dimension} & \textbf{Liquidity-Saving Mechanisms (LSMs)} & \textbf{Cycles Protocol} \\
        \hline
        \textbf{Institutional Scope} &
        Central bank RTGS systems; access limited to licensed banks and payment institutions. &
        Open, neutral settlement layer across firms and market infrastructures. \\
        \hline
        \textbf{Settlement Mechanism} &
        Queuing and offsetting of homogeneous payment instructions via netting. &
        Atomic multilateral \textit{setoff} across heterogeneous obligations via cycle detection. \\
        \hline
        \textbf{Settlement Asset} &
        Central bank reserves only. &
        Any verifiable digital asset or tokenized claim denominated in a common unit of account. \\
        \hline
        \textbf{Network Topology} &
        %Centralized, hub-and-spoke structure via central counterparty. &
        Centralized payment-system operator / central bank &
        Decentralized graph of balance sheets without central counterparty. \\
        \hline
        \textbf{Risk Model} &
        %Risk concentration via central counterparty and collateral management. &
        Payment-system liquidity optimization &
        Risk reduction via balance-sheet compression without collateral. \\
        \hline
        \textbf{Liquidity Efficiency} &
        Reduces intraday reserve demand within banking system. &
        Reduces liquidity requirements within commercial and financial economy. \\
        \hline
        \textbf{Regulatory Boundary} &
        Payment system oversight under central bank jurisdiction. &
        Multiple regulatory domains (payments, securities, trade credit). \\
        \hline
        \textbf{Privacy and Transparency} &
        Full visibility to operator; limited confidentiality. &
        Privacy-preserving through zero-knowledge proofs and Trusted Execution Environments. \\
        \hline
    \end{tabularx}
    \caption{Comparison between Liquidity-Saving Mechanisms (LSMs) in RTGS systems and the Cycles Protocol. The Cycles design extends LSM principles beyond interbank payments to encompass multi-asset and real-economy settlement.}
    \label{tab:lsm_comparison}
\end{table}

By internalizing trade credit liquidity, the protocol effectively extends the central bank’s liquidity-saving logic to the broader economy without direct balance sheet exposure.

\section{Discussion and Implications}

The clearing paradigm presented in this paper, which unifies financial market and real-economy obligations, has implications for the design of market infrastructure. Although trade credit integration may initially seem peripheral to securities exchanges, it provides a concrete mechanism for expanding the set of obligations that can be jointly cleared and for increasing settlement liquidity through multilateral setoff. The historical separation between these liquidity pools reflects coordination and institutional constraints rather than a lack of economic linkage. The Cycles Protocol can be viewed as a neutral settlement layer that reduces coordination complexity by allowing heterogeneous obligations to be represented and discharged on a single clearing graph. This proposal finds precedent in Latin American markets, where invoices have long been treated as tradable securities, and in Slovenia, where trade credit clearing has operated at the national scale for decades,
suggesting that the primary barrier in other jurisdictions has been less conceptual and more a result of institutional path dependency. Unlike the bank-dominated settlement infrastructures common elsewhere, Slovenia’s model evolved from a pre-90s non-bank payment system, which fostered a fundamentally different operational perspective on how obligations could be cleared directly.

By providing a decentralized method for balance sheet compression, the protocol may reduce some forms of operational and liquidity dependency on 
CCPs by lowering gross settlement liquidity demands in interconnected obligation graphs. This speaks to a core dilemma: while CCPs are efficient, they also create a central risk nexus \cite{gaffeo2019economics}. In that sense, the Cycles Protocol suggests a path to achieve comparable setoff efficiency without increasing risk concentration via novation, while the broader systemic-risk implications remain to be validated in realistic deployments and pilot studies.

The implications of this integration with the Cycles Protocol vary between stakeholders, but converge on a single principle: \textbf{liquidity efficiency through balance-sheet connectivity}.

\subsubsection*{For Exchanges and Market Operators}

The most direct implication is the potential for significant market expansion and product innovation. 
By treating trade credit (AR/AP) as a liquidity source, the protocol integrates real-economy obligations into post-trade operations.
This opens the door to creating regulated \textbf{trade credit marketplaces}, where corporate receivables can be securitized. This type of offering would not only create a new valuable asset class, but would also facilitate the \textbf{inclusion of Small and Medium Enterprises (SMEs)} into the financial ecosystem on an unprecedented scale. For SMEs, whose primary asset is often their receivables, this transforms a static balance sheet item into a dynamic source of liquidity, lowering their cost of capital and enabling greater participation in the formal economy. This, in turn, feeds back into the core market, deepening the overall liquidity pool available for securities settlement, and creating a more robust and diverse ecosystem.

\subsubsection*{For Clearing Systems and Central Counterparties}

For existing clearing infrastructures, including CCPs, our framework offers a path toward greater settlement-liquidity efficiency and operational flexibility. The protocol is not designed to replace CCPs, but to complement them by providing a mechanism for settling complex, multilateral, and cross-asset payment obligations that are ill-suited to the traditional novation model.
Critically, Cycles does not interfere with CCPs' core credit-risk management functions. Margining requirements, default fund contributions, and loss mutualization procedures remain entirely within the CCP domain.
As our stylized model demonstrates, the protocol can reduce gross settlement liquidity requirements by settling interlocking positions that span multiple CCPs, thereby lowering the cash that clearing members must source and transfer to meet a given set of payment obligations. Whether and how this translates into lower collateral usage or lower margin requirements depends on specific CCP margin models and rulebooks and is therefore outside the model's scope. Likewise, potential effects on procyclicality and systemic resilience remain hypotheses for pilot testing. By lowering gross settlement cash needs in fragmented arrangements, the protocol \textbf{may reduce some operational and liquidity dependencies on CCP silos}, while leaving CCP concentration unchanged with respect to novation, loss mutualization, and default management.

\subsubsection*{For Regulators and Supervisors}

This new paradigm naturally introduces novel challenges and considerations for regulatory oversight, yet its design incorporates solutions to the most critical concerns. The primary task will be to establish frameworks for \textbf{credit oversight} within this integrated ecosystem. The protocol helps address this at the source through its logic of \textbf{double ascertainment}: before any trade receivable can be used for settlement, it must be digitally confirmed by both the creditor and debtor, reducing the entry of erroneous or fraudulent obligations into the system. This transforms informal credit into a verified asset class that is ready for settlement and provides a robust first line of defense against bad-faith actors.

Supervisors will also need to analyze potential new \textbf{systemic risk channels}. Here, the protocol offers a fundamental departure from the CCP model by operating on a principle of \textbf{risk reduction, not risk redistribution}. Unlike a CCP, which mutualizes member risk through a default fund, the protocol only executes fully solvent, pre-agreed cycles. 
A default affects only the defaulting party’s obligation chain, with damage contained near the source. Cycles does not execute settlement flows for parties that fail to provide required liquidity, limiting impact to direct relationships rather than spreading losses system-wide. Losses are not mutualized: unaffected participants do not bear others’ defaults, unlike in CCP structures where members share losses via default funds.

Finally, the inherent transparency of a DLT-based system offers a powerful new supervisory tool. It could provide regulators with a real-time, system-wide view of liquidity, leverage, and emerging risk concentrations that is difficult to achieve in today's fragmented and opaque financial plumbing.

\subsubsection*{Failure Modes and Stress Considerations}

Linking previously separate obligation networks expands the clearing graph and increases the number of offset paths, which can lower the gross settlement cash requirements. However, a larger clearing space also means that \emph{the quality and availability of trade-credit edges matter}. If participants come to rely on receivables and payables as a material settlement-liquidity source, then stress in the real economy can reduce clearing capacity. For example, if receivables become disputed, delayed, or ineligible, liquidity savings can evaporate precisely when funding is scarce. The implication is complementarity, not substitution. CCPs continue to manage counterparty credit risk via novation, margining, and default waterfalls, while Cycles targets \emph{settlement liquidity} by canceling solvent cycles across a broader set of obligations. A win--win outcome therefore depends on conservative eligibility rules, concentration limits, and operational fallbacks that reduce the chance that trade credit becomes a point of settlement failure.

\section{Conclusion}

This paper has introduced a new paradigm for post-trade clearing, moving beyond the centralized, novation-based model of CCPs toward a decentralized network for direct balance sheet compression. We have argued that a primary obstacle to settlement efficiency is the fragmentation of liquidity across institutional and economic silos. The vast reservoir of working capital locked in trade credit has remained separate from the world of securities settlement primarily because of the coordination problem involved in integrating them. Our central contribution is the Cycles Protocol: a theoretical and practical framework that addresses this fragmentation. By representing both past-due obligations and future-facing commitments on a single, unified clearing graph, the protocol discovers and atomically executes multilateral settlement cycles. As demonstrated, this approach can reduce gross settlement cash needs in inter-CCP fragmented settings and mobilize trade-credit obligation networks as an additional source of settlement liquidity. Rather than replacing CCP credit-risk management, the protocol targets settlement-liquidity efficiency through direct cancellation of interwoven balance-sheet claims, with broader systemic effects depending on design choices and stress behavior.

The implications of this shift are significant. For market operators, this framework offers a concrete path to deepen settlement liquidity, create new products such as trade credit marketplaces, and promote greater inclusion of SMEs in the formal financial system. For the global financial system, it provides a potential mechanism to reduce operational and liquidity reliance on siloed settlement channels by reducing gross settlement cash requirements in fragmented clearing arrangements. The next step is to move from theory to practice through \textbf{pilot studies with market participants} that evaluate operational feasibility, governance, and legal enforceability, and that measure liquidity and risk outcomes under realistic constraints.

\bibliographystyle{plain}
\bibliography{references}

\end{document}